\documentclass[aps,prd,showpacs,floatfix,notitlepage,nofootinbib,twocolumn,a4paper]{revtex4}
\usepackage{amsfonts,amsmath,units,wasysym,epsfig,graphicx,verbatim,color,subfigure,graphicx,bm,mathrsfs,lipsum,hyperref,cleveref}
\usepackage{booktabs}
\usepackage[normalem]{ulem}  % \sout{old text} for strikeout
\definecolor{dgreen}{RGB}{0, 204, 0}

\newcommand{\be}{\begin{equation}}
\newcommand{\ee}{\end{equation}}
\newcommand{\bea}{\begin{eqnarray}}
\newcommand{\eea}{\end{eqnarray}}

\DeclareUnicodeCharacter{2212}{-}
\usepackage[none]{hyphenat}

\usepackage{cellspace}
\begin{document}

% \setcounter{page}{1}
% \vspace*{0.3 true in}
%\title{Impacts of PSR J0614−3329 on the Nuclear Matter Parameters Constraints}
\title{Bayesian Analysis of the Neutron Star EoS and Model Comparison: Insights from PSR J0437+4715, PSR J0614+3329, and Other Multi-Physics Data}

\author{\href{https://orcid.org/0000-0003-3308-2615}Sk Md Adil Imam$^{1}$\includegraphics[scale=0.06]{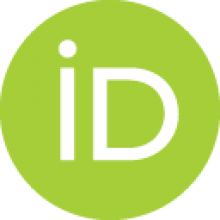}}
\email{adil.imam@unab.cl}

\author{\href{https://orcid.org/0000-0003-0103-5590}N. K. Patra$^{2}$\includegraphics[scale=0.06]{Orcid-ID.png}}
\email{nareshkumarpatra3@gmail.com}

\affiliation{$^1$Instituto de Astrof\'isica, Departamento de F\'isica y Astronom\'ia, Universidad Andr\'es Bello,Santiago, Chile}

\affiliation{$^2$School of Science and Engineering, The Chinese University of Hong Kong, Shenzhen (CUHKShenzhen), Guangdong, 518172, China}

\date{\today}

\begin{abstract} 
We perform a comprehensive Bayesian analysis to constrain the neutron star (NS) equation of state (EoS) using a wide range of terrestrial and astrophysical data. The terrestrial inputs include quantities related to symmetric nuclear matter (SNM) and symmetry energy up to two times saturation density ($\rho_0 \simeq$ 0.16 fm$^{-3}$), derived from finite nuclei and HIC. The astrophysical constraints incorporate NS radii and tidal deformabilities from recent NICER observations and GW170817, respectively. We consider five different EoS models: Taylor, $n/3$, Skyrme, RMF, and sound speed (CS), and analyze them by sequentially updating the priors with (i) $\chi$EFT-based pure neutron matter, (ii) terrestrial, empirical and earlier astrophysical data, (iii) case (ii) including NICER radii of PSR J0437+4715 and J0614+3329, (iv) all data combined, and (v) excluding empirical nuclear inputs. We also perform Bayesian model comparison which favors the Skyrme model under all combined data [scenario (iv)], yielding tight constraints on symmetry energy parameters: $L_0 = 56 \pm 3$~MeV, $K_{\mathrm{sym}0} = -132 \pm 15$~MeV and also on SNM parameters: $K_0 = 265 \pm 12$~MeV and $Q_0 = -366 \pm 43$~MeV. The mass-radius and mass-tidal deformability posterior distributions are also well constrained. The radius and tidal deformability of a $1.4\,M_\odot$ neutron star are found to be $R_{1.4} = 11.85 \pm 0.11$~km and $\Lambda_{1.4} = 354 \pm 25$, respectively.

\end{abstract}

\maketitle
\section{Introduction}

Neutron stars (NSs) are among the most compact objects in the universe and serve as natural laboratories to study cold and dense matter under extreme conditions~\cite{Glendenning1997, Haensel:2007yy, FiorellaBurgio:2018dga, Rezzolla:2018jee}. The equation of state (EoS) of dense matter plays a central role in understanding the structure and behavior of neutron stars, linking nuclear physics with astrophysics. Neutron star matter is in a state of $\beta-$equilibrium and charge neutrality, making it highly neutron-rich and asymmetric. The nuclear part of the EoS is usually described through symmetric nuclear matter (SNM) and the density-dependent symmetry energy, both of which are expressed using nuclear matter parameters (NMPs) such as the energy per nucleon for SNM, symmetry energy, and their derivatives at nuclear saturation density ($\rho_0 \simeq$ 0.16 fm$^{-3}$). The lower-order NMPs, important for the EoS at low densities, are typically constrained by laboratory experiments and nuclear models based on finite nuclei. In contrast, higher-order NMPs, which influence the high-density behavior relevant to neutron star cores, are more uncertain and often inferred from astrophysical observations like the maximum mass, radius, and tidal deformability of neutron stars. Due to the limited availability of experimental data, these observational constraints are widely used to narrow down the possible range of NMPs to a smaller subspace.

In recent years, observational astronomy has provided powerful tools to constrain the EoS of neutron star matter~\cite{Abbott:2017vwq, Abbot2018, Abbot2019, Miller:2019cac, Riley:2019yda, Miller:2021qha, Riley:2021pdl, Lattimer:2012nd, Hebeler:2013nza, Lattimer:2021emm, Ferreira:2021pni, Zhang:2018vrx, Malik:2018zcf}. One of the most important developments is gravitational wave (GW) astronomy, which studies the signals emitted during binary neutron star mergers. The first such event, GW170817, detected by the 
advanced LIGO~\cite{LIGOScientific:2014pky}  and Virgo detectors~\cite{VIRGO:2014yos}, provided the first measurement of the tidal deformability of neutron stars—a quantity that directly reflects the stiffness of the EoS~\cite{ Malik:2018zcf, Chatziioannou:2020pqz, Baiotti:2019sew, Li:2021thg, Lim:2019som, Raithel:2019uzi, Fattoyev:2017jql, Forbes:2019xaz, Landry:2020vaw, Piekarewicz:2018sgy, Biswas:2020puz, Abbott:2020aai, Thi:2021hai, Patra:2022lds, Sorensen:2023zkk, Soma:2023rmq, Gao:2025vdc, Tang:2025xib}. A second likely binary neutron star merger, GW190425, was also observed. These events have triggered many theoretical studies aimed at narrowing the possible forms of the EoS. Future observations from LIGO-Virgo-KAGRA and next-generation detectors like the Einstein Telescope\cite{Punturo:2010zz}, and Cosmic Explorer~\cite{Reitze:2019iox} are expected to detect many more such signals. Alongside gravitational waves, precise measurements of neutron star masses and radii by the Neutron Star Interior Composition Explorer (NICER) mission have added complementary constraints~\cite{Watts2016, Psaltis2014}. For example, NICER has measured the mass and radius of pulsars like PSR J0437+4715~\cite{Choudhury:2024xbk}, PSR J0614+3329~\cite{Mauviard:2025dmd}, PSR J0030+0451~\cite{Riley:2019yda,Miller:2019cac}and PSR J0740+6620~\cite{Riley:2021pdl,Miller:2021qha}  with high accuracy. The latter is a particularly massive pulsar, and such heavy neutron stars suggest that the EoS must be stiff enough to support masses greater than 2 $M_\odot$. Typically, data from nuclear physics experiments and astrophysical observations are employed to constrain the EoS using a Bayesian framework \cite{Tews2012,Hebeler:2013nza,LeFevre2015,Russotto2016,Lynn2016,Monitor2017,Coughlin2017,Drischler2017,Zhang2018,Drischler2020,Dietrich2020,Raaijmakers2020,Huth2021,Biswas2021,Tsang:2023vhh, Imam:2021dbe,Malik2024b,Imam:2024xqg,Tewari:2024qit}. With more data, it is becoming increasingly possible to use these observations to constrain the nuclear matter parameters which underlie the construction of the EoS of neutron star matter.

The present lower bound on $M_{\rm max}$ suggests that the central density of a neutron star with canonical mass $1.4,M_\odot$ may lie in the range $\sim 2$–$3,\rho_0$~\cite{Hebeler:2013nza, Li:2005sr}, indicating that the behavior of the EoS near the nuclear saturation density $\rho_0$ plays a key role in shaping the structure of such stars. Consequently, several studies have investigated the correlations between neutron star observables—such as mass, radius, and tidal deformability—and the nuclear matter parameters, particularly those governing the density dependence of the symmetry energy~\cite{Alam:2016cli, Carson:2018xri, Malik:2018zcf, Tsang:2019vxn, Patra:2023jbz, Imam:2023ngm, Patra:2023jvv}. While lower-order NMPs like the symmetry energy $J$ and its slope $L$ at $\rho_0$ are relatively better constrained from experimental data on finite nuclei, higher-order NMPs—such as the curvature $K_{\rm sym}$ and skewness parameters—remain largely uncertain. These higher-order parameters, which become increasingly important at suprasaturation densities relevant to the neutron star core, continue to pose significant challenges for precise determination. Bayesian statistical frameworks have become common tools for exploring the posterior distributions of NMPs and neutron star observables, though the resulting correlations are sensitive to model assumptions and the nature of the input data~\cite{Malik:2018zcf, Reed:2021nqk}. The limitations of current observational and experimental data lead to wide uncertainties in the extracted NMPs, especially since terrestrial experiments probe the EoS indirectly, and converting such constraints into NMPs introduces further model dependence. For instance, while the slope parameter $L$ is typically quoted around $60 \pm 20$ MeV, recent analyses report larger values, highlighting discrepancies among different approaches~\cite{Tsang:2020lmb, Reed:2021nqk}. Meanwhile, recent efforts employing hundreds of calibrated mean-field models—both relativistic and nonrelativistic have shown that NMPs constrained by finite nuclear properties lead to stronger and more reliable correlations with neutron star observables~\cite{Carlson:2022nfb}. In contrast, models using broad, uncorrelated priors yield weak or ambiguous correlations, underscoring the necessity of integrating astrophysical observations with terrestrial nuclear constraints. Despite these advances, higher-order NMPs remain highly uncertain, and their imprecise values significantly affect predictions of neutron star properties at high densities. Therefore, continued efforts are essential to better constrain the higher-order NMPs, as they are crucial for developing a unified and accurate EoS for dense matter in neutron stars.

In this work, we aim to constrain the dense matter EoS within a Bayesian framework using five different EoS models: Taylor, $n/3$, Skyrme, relativistic mean field (RMF), and speed-of-sound (CS). Our analysis focuses on extracting key nuclear matter properties, including the symmetric nuclear matter energy, the density-dependent symmetry energy, and their density derivatives, by systematically incorporating diverse and complementary constraints. In particular, we include constraints from heavy ion collisions (HICs), empirical nuclear inputs, $\chi$EFT calculations of pure neutron matter (PNM), and astrophysical observations of neutron stars. 
To assess their impact, the data are organized into five sets as \textbf{Set1}: $\chi$EFT PNM, \textbf{Set2}: terrestrial data, empirical nuclear inputs, and earlier astrophysical observations, \textbf{Set3}: Set2 plus new NICER radii measurements of PSR~J0437+4715 and PSR~J0614+3329, \textbf{Set4}: all data combined, i.e., Set1 and Set3, and \textbf{Set5}:  Set4 excluding empirical nuclear inputs. We investigate the influence of each dataset on the posterior distributions of NMPs and neutron star properties. Furthermore, using the Bayes factor, we perform a model comparison analysis to identify the most favored EoS model. 
Through this comprehensive approach, we seek to bridge terrestrial nuclear experiments with multimessenger astrophysical observations, thereby providing tighter and more robust constraints on the nuclear EoS.

The paper is organized as follows, Sec.~\ref{EoS} provides a brief overview of the various EoS models employed for neutron star matter. The Bayesian analysis framework is described in Sec.~\ref{Bayes}. In Sec.~\ref{data}, we summarize the observational and experimental data utilized in this study. The results and their implications are presented and discussed in Sec.~\ref{PD}. Finally, conclusions are drawn in Sec.~\ref{Conclusion}.
%%%%%%%%%%%%%%%%%%%%%%%%%%%%%%%%%%%%%%%%%%
\section{EQUATION OF STATE MODELS} 
\label{EoS}
We discuss in brief the construction of the EoS using the five different models. The total energy per nucleon \( \varepsilon(\rho, \delta) \) at a given total nucleon density $\rho$ and asymmetry parameter $\delta$ can be decomposed into the energy per nucleon for
the SNM, e$(\rho, 0)$ and the density-dependent symmetry energy,
J($\rho$) in the parabolic approximation as
\bea
\varepsilon(\rho, \delta) &=&  e(\rho,0)+ J(\rho)\delta^2
+..., \label{eq:EoS}
\eea
where $\delta = \frac{\rho_n -\rho_p}{\rho}$
is determined  using the $\beta$-equilibrium and the charge neutrality conditions.

The energy per particle for SNM \(e(\rho, 0)\) and symmetry energy \(J(\rho)\) can be evaluated for a given EoS model, using nuclear matter parameters,
a$_n$ and b$_n$, which are defined as
\begin{equation}
a_n = (3\rho_0)^n \frac{d^n e(\rho, 0)}{d\rho^n} \Bigg|_{\rho = \rho_0}
\end{equation}

where,
\(a_0 = e_0\) (binding energy),
\(a_1 = 0\),
\(a_2 = K_0\) ( incompressibility coefficient),
\(a_3 = Q_0\) (skewness),
\(a_4 = Z_0\) (kurtosis), and

\begin{equation}
b_n = (3\rho_0)^n \frac{d^n J(\rho, 0)}{d\rho^n} \Bigg|_{\rho = \rho_0}
\end{equation}

where
\(b_0 = J_0\) (symmetry energy coefficient),
\(b_1 = L_0\) (slope parameter),
\(b_2 = K_{\text{sym0}}\) (curvature parameter),
\(b_3 = Q_{\text{sym0}}\) (skewness),
\(b_4 = Z_{\text{sym0}}\) (kurtosis).

\subsection{Taylor Expansion}\label{T}
 The energy per nucleon for the symmetric nuclear matter, e$(\rho,0)$ and the density-dependent symmetry energy, J$(\rho)$ are expanded around  $\rho_0$ using individual nuclear matter parameters as~\cite{Chen:2005ti,Chen:2009wv,Newton:2014iha,Margueron:2017eqc,Margueron:2018eob}
\begin{equation}
e(\rho, 0) = \sum_{n=0}^{4} \frac{a_n}{n!} x^n\label{TE}
\end{equation}

\begin{equation}
J(\rho) = \sum_{n=0}^{4} \frac{b_n}{n!} x^n\label{TJ}
\end{equation}

where \( x = \frac{\rho - \rho_0}{3\rho_0} \).

\subsection{$\frac{n}{3}$ Expansion}\label{n3}
Alternative expansion of \( \varepsilon(\rho, \delta) \) can be obtained by expanding \( e(\rho, 0) \) and \( J(\rho) \) as~\cite{Lattimer:2015nhk, Gil:2016ryz}

\begin{equation}
e(\rho, 0) = \sum_{n=2}^{6} a_{n-2}^{\prime} y^{n/3}\label{n3E}
\end{equation}

\begin{equation}
J(\rho) = \sum_{n=2}^{6} b_{n-2}^{\prime} y^{n/3}\label{n3J}
\end{equation}
where \( y = \frac{\rho}{\rho_0} \). The coefficients $a_{n-2}^{\prime}$ and $b_{n-2}^{\prime}$ are expressed in terms of NMPs as
\[
\begin{bmatrix}
a_0' \\
a_1' \\
a_2' \\
a_3' \\
a_4'
\end{bmatrix}
=
\frac{1}{24}
\begin{bmatrix}
360 & 20 & 0 & 1 \\
-960 & -56 & -4 & -4 \\
1080 & 60 & 12 & 6 \\
-576 & -32 & -12 & -4 \\
120 & 8 & 4 & 1
\end{bmatrix}
\begin{bmatrix}
e_0 \\
0   \\
K_0 \\
Q_0 \\
Z_0
\end{bmatrix}
\]

\[
\begin{bmatrix}
b_0' \\
b_1' \\
b_2' \\
b_3' \\
b_4'
\end{bmatrix}
=
\frac{1}{24}
\begin{bmatrix}
360 & -120 & 20 & 0 & 1 \\
-960 & 328 & -56 & -4 & -4 \\
1080 & -360 & 60 & 12 & 6 \\
-576 & 192 & -32 & -12 & -4 \\
120 & -40 & 8 & 4 & 1
\end{bmatrix}
\begin{bmatrix}
J_0 \\
L_0 \\
K_{\text{sym},0} \\
Q_{\text{sym},0} \\
Z_{\text{sym},0}
\end{bmatrix}
\]

\subsection{Skyrme}\label{Skyrme}
The Hamiltonian density for uniform nuclear matter using a Skyrme-type effective interaction is given by
\begin{equation}
\label{Hden}
\mathcal{H} = \mathcal{K} + \mathcal{H}_0 + \mathcal{H}_3 + \mathcal{H}_{\text{eff}},
\end{equation}
where $\mathcal{K} = \frac{\hbar^2}{2m}\tau$ represents the kinetic energy term, $\mathcal{H}_0$ is the zero-range interaction, $\mathcal{H}_3$ the density-dependent term, and $\mathcal{H}_{\text{eff}}$ accounts for effective mass corrections.

The contributions from the Skyrme interaction are
\begin{equation}
\mathcal{H}_0 = \frac{1}{4}t_0\left[(2+x_0)\rho^2 - (2x_0+1)(\rho_p^2 + \rho_n^2)\right],
\end{equation}
\begin{equation}
\mathcal{H}_3 = \frac{1}{24}t_3\rho^\sigma\left[(2+x_3)\rho^2 - (2x_3+1)(\rho_p^2 + \rho_n^2)\right],
\end{equation}
\begin{align}
\mathcal{H}_{\text{eff}} = &\ \frac{1}{8}\left[t_1(2+x_1) + t_2(2+x_2)\right]\tau\rho \nonumber \\
& + \frac{1}{8}\left[t_2(2x_2+1) - t_1(2x_1+1)\right](\tau_p\rho_p + \tau_n\rho_n).
\end{align}

Here, $\rho = \rho_p + \rho_n$ and $\tau = \tau_p + \tau_n$ are the total nucleon density and kinetic energy density, respectively, with $p$ and $n$ denoting protons and neutrons. The kinetic energy density for each component is
\begin{align}
\tau_q &= \frac{\gamma}{(2\pi)^3}\int_{0}^{k_{f_q}}k^2\,d^3k \nonumber \\
&= \frac{3}{5}(3\pi^2)^{2/3}\rho_q^{5/3},
\end{align}
where $q = p,n$ and $\gamma = 2$ denotes spin degeneracy.

For symmetric nuclear matter, where $\rho_p = \rho_n = \frac{\rho}{2}$, the energy per nucleon is given by

\begin{equation}
e(\rho, 0) = c_1 \cdot \rho^{\frac{2}{3}} + \frac{3}{8} \cdot t_0 \cdot \rho + \frac{3}{10} \cdot \theta_{\text{\rm SNM}}  \cdot \rho^{\frac{5}{3}} + \frac{1}{16} \cdot t_3 \cdot \rho^{\sigma+1}\label{esnm_skyrme}
\end{equation}
where \[c_1 = \frac{3\hbar^2}{10m}(\frac{3\pi^2}{2})^\frac{2}{3}\]

\[
\sigma = \frac{1}{3} \cdot \frac{K_0 + Q_0 - 6 c_1 \rho_0^{\frac{2}{3}}}{15 e_0 + K_0 - 3 c_1 \rho_0^{\frac{2}{3}}} - 1
\]

\[
t_3 = \frac{16 (15 e_0 + K_0 - 3 c_1 \rho_0^{\frac{2}{3}})}{3 \sigma (3 \sigma - 2) \rho_0^{\sigma + 1}}
\]

\[
\theta_{\text{\rm SNM}} = \frac{5 \left( c_1\rho_0^{\frac{2}{3}} / 3 - e_0 - \frac{\sigma t_3}{16} \rho_0^{\sigma + 1} \right)}{ \rho_0^{\frac{5}{3}}}
\]

\[
t_0 = \left( e_0 - c_1\rho_0^{\frac{2}{3}} - \frac{3}{10} \theta_{\text{snm}} \left( \frac{3 \pi^2}{2} \right)^{\frac{2}{3}} \rho_0^{\frac{5}{3}} - \frac{t_3}{16} \rho_0^{\sigma + 1} \right)\left(\frac{ 8}{3 \rho_0}\right)
\]

\begin{equation}
J(\rho) = c_2 \cdot \rho^{\frac{2}{3}} + a_1 \cdot \rho + a_{\text{sym}} \cdot \rho^{\frac{5}{3}} + a_\sigma \cdot \rho^{\sigma+1}\label{esym_skyrme}
\end{equation}

where
\[
c_2=\frac{5}{9}c_1
\]
\[
a_{\sigma} = \frac{K_{\text{sym}0} - 5 \left( L_0 - 3 J_0 \right) - 3 c_2 \rho_0^{\frac{2}{3}}}{\left( 3 \sigma \right) \left( 3 \sigma - 2 \right) \rho_0^{\sigma + 1}}
\]

\[
a_{\text{sym}} = \frac{\left( L_0 - 3 J_0 \right) - 3 \sigma a_{\sigma} \rho_0^{\sigma + 1} + c_2 \rho_0^{\frac{2}{3}}}{2 \rho_0^{\frac{5}{3}}}
\]

\[
a_1 = \frac{J_0}{\rho_0} - c_2 \rho_0^{-\frac{1}{3}} - a_{\text{sym}} \rho_0^{\frac{2}{3}} - a_{\sigma} \rho_0^{\sigma}
\]

\subsection{RMF}\label{RMF}

We generate a set of equations of state within the framework of relativistic mean field theory, where nucleon–nucleon interactions are mediated by meson exchange: the scalar–isoscalar $\sigma$, vector–isoscalar $\omega$, and vector–isovector $\rho$. In this work, we adopt the general nonlinear finite-range RMF model~\cite{Dutra:2014qga, Todd-Rutel:2005yzo}, represented by the following Lagrangian density

\begin{eqnarray}
\mathcal{L} = \mathcal{L}_{\rm N} + \mathcal{L}_{\rm M} +
\mathcal{L}_{\rm NL},
\label{eq-Lagrangian}
\end{eqnarray}

where
\begin{align}
\mathcal{L}_{\rm N} &= \overline{\psi}(i\gamma^\mu\partial_\mu - m)\psi 
+ g_\sigma\sigma\overline{\psi}\psi 
- g_\omega\overline{\psi}\gamma^\mu\omega_\mu\psi \nonumber \\ 
&- \frac{g_\rho}{2}\overline{\psi}\gamma^\mu\vec{\rho}_\mu\vec{\tau}\psi,\nonumber
\end{align}
\begin{align}
\mathcal{L}_{\rm M} &= \frac{1}{2}(\partial^\mu \sigma \partial_\mu \sigma - m^2_\sigma\sigma^2)
-\frac{1}{4}F^{\mu\nu}F_{\mu\nu} \nonumber\\
&+ \frac{1}{2}m^2_\omega\omega_\mu\omega^\mu 
-\frac{1}{4}\vec{B}^{\mu\nu}\vec{B}_{\mu\nu} 
+ \frac{1}{2}m^2_\rho\vec{\rho}_\mu\vec{\rho}^\mu,\nonumber
\end{align}
and
\begin{align}
\mathcal{L}_{\rm NL} &= - \frac{g_2}{3}\sigma^3 - \frac{g_3}{4}\sigma^4 + \frac{c_3}{4}(g_\omega^2\omega_\mu\omega^\mu)^2\nonumber\\
&+ \frac{1}{2}{\alpha_3}'g_\omega^2 g_\rho^2\omega_\mu\omega^\mu
\vec{\rho}_\mu\vec{\rho}^\mu.\nonumber
\end{align}

In this Lagrangian density, $\mathcal{L}_{\rm N}$ denotes the nucleon kinetic term, while $\mathcal{L}_{\rm M}$ contains the free and self-interaction terms of the mesons $j=\sigma,\omega,\rho$. The nonlinear term $\mathcal{L}_{\rm NL}$ accounts for crossed interactions among the meson fields. The antisymmetric field tensors are defined as $F_{\mu\nu}=\partial_\nu\omega_\mu-\partial_\mu\omega_\nu$ and $\vec{B}_{\mu\nu}=\partial_\nu\vec{\rho}_\mu-\partial_\mu\vec{\rho}_\nu - g_\rho (\vec{\rho}_\mu \times \vec{\rho}_\nu)$. The nucleon mass is fixed at $m=939$~MeV, and the meson masses at $m_\sigma=491.5$~MeV, $m_\omega=782.5$~MeV, and $m_\rho=763$~MeV. The nucleon–meson couplings are denoted by $g_\sigma$, $g_\omega$, and $g_\rho$, respectively. 

The parameters $g_2$, $g_3$, $c_3$, and $\alpha_3^\prime$ determine the strength of the nonlinear terms: $g_2$ and $g_3$ govern the nuclear matter incompressibility at saturation density $\rho_0$~\cite{Boguta:1977xi}, $c_3$ controls the high-density stiffness of the EoS (larger $c_3$ yielding a softer EoS), and $\alpha_3^\prime$ affects the density dependence of the symmetry energy. The influence of these nonlinear terms is evident in the meson field equations, such as

\begin{align}
m^2_\sigma\sigma &= g_\sigma\rho_s - g_2\sigma^2 - g_3\sigma^3 \mbox{,}\quad 
\label{eq-sigma-field}\\
m_\omega^2\omega_0 &= g_\omega\rho - c_3.g_\omega(g_\omega \omega_0)^3 - {\alpha_3}'g_\omega^2 g_\rho^2\bar{\rho}_{0(3)}^2\omega_0, 
\label{eq-omega-field}\\
m_\rho^2\bar{\rho}_{0(3)} &= \frac{g_\rho}{2}\rho_3 
-{\alpha_3}'g_\omega^2 g_\rho^2\bar{\rho}_{0(3)}\omega_0^2, 
\label{eq-rho-field}
\end{align}

where $\rho_s$ and $\rho_{s3}$ are scalar and vector densities, respectively.

Through the Lagrangian density in Eq.~(\ref{eq-Lagrangian}), we can obtain the energy per particle for the symmetric nuclear matter ($M_p^*=M_n^*=M^*=M-g_\sigma\sigma$) as 
\begin{eqnarray}
e(\rho, 0) &=& \frac{1}{2}m^2_\sigma\sigma^2 
+ \frac{g_2}{3}\sigma^3 + \frac{g_3}{4}\sigma^4 - \frac{1}{2}m^2_\omega\omega_0^2 
- \frac{c_3}{4}(g_\omega^2\omega_0^2)^2 \nonumber \\
&-& \frac{1}{2}m^2_\rho\bar{\rho}_{0(3)}^2
+g_\omega\omega_0\rho+\frac{g_\rho}{2}\bar{\rho}_{0(3)}\rho_3
\nonumber \\
&-& \frac{1}{2}{\alpha_3}'g_\omega^2 g_\rho^2\omega_0^2\bar{\rho}_{0(3)}^2
+ \mathcal{E}_{\mbox{\tiny kin}}^p + \mathcal{E}_{\mbox{\tiny kin}}^n,
\label{eq-esnm}
\end{eqnarray}

where 
\begin{eqnarray}
\mathcal{E}_{\mbox{\tiny kin}}^{p,n}&=&\frac{\gamma}{2\pi^2}\int_0^{{k_F}_{p,n}}k^2
(k^2+M^{*2}_{p,n})^{1/2}dk 
\nonumber \\
&=& \frac{3}{4}{E_{F}}_{p,n}\rho_{p,n} + \frac{1}{4}M^{\ast}_{p,n}{\rho_{s}}_{p,n},\label{eq-ekin}
\end{eqnarray}
and
\begin{equation}
{E_{F}}_{p,n} = \sqrt{{k_{F}}_{p,n}^{2}+(M^{\ast}_{p,n})^{2}}.
\end{equation} 
The symmetry energy, ${J}(\rho)$, can be calculated as 
\begin{align}
{J}(\rho) &= \frac{1}{8}\frac{\partial^2(e/\rho)}{\partial
y^2}\bigg|_{\rho,y=1/2}, \nonumber\\
&= \frac{k_F^2}{6E_F^*}
+ \frac{g_\rho^2}{8{m_\rho^*}^2}\rho.\label{eq-esym}
\end{align}
where $E_F^*=(k_F^2+{M^*}^2)^{1/2}$ is the fermi momentum energy. 

\begin{figure}
    \centering
    \includegraphics[width=0.9\linewidth]{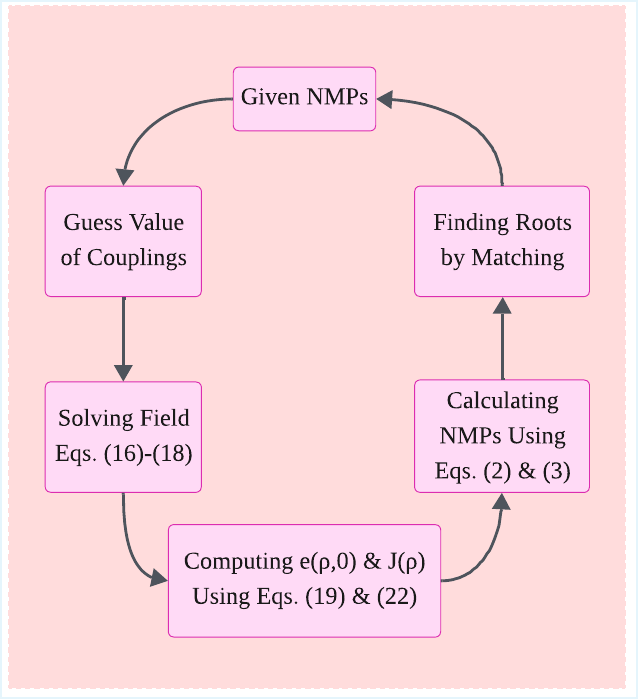}
    \caption{A flowchart for translating the coupling constants into the NMPs within the RMF framework, enabling the use of prior distributions on NMPs rather than on the couplings.}
    \label{fig1}
\end{figure}

Figure~\ref{fig1} illustrates the mapping of coupling constants to nuclear matter parameters within the RMF framework. Since imposing priors directly on couplings is not straightforward, we assign priors to the NMPs. Starting from these priors, a trial set of couplings is provided, and the meson field equations are solved to obtain the energy of symmetric and asymmetric matter, from which the NMPs are recalculated. The couplings are iteratively adjusted until the recalculated NMPs converge to the assigned values. This root-finding procedure ensures consistency between the NMP priors and RMF couplings, thereby yielding the corresponding EoS.

Figure~\ref{fig1} illustrates the procedure for translating coupling constants into nuclear matter parameters within the RMF framework. Since it is not straightforward to impose priors directly on the couplings, we instead assign priors to the NMPs. Starting from the chosen NMPs, a set of trial coupling constants is provided as input. The field equations are then solved to obtain the energy of symmetric and asymmetric nuclear matter, from which the NMPs are recalculated. The procedure iterates by adjusting the couplings until the recalculated NMPs converge to the initially assigned values. This root-finding approach ensures consistency between the chosen NMP priors and the RMF couplings, ultimately yielding the corresponding EoS.

\begin{table}[h]
\caption{\label{tab1:prior}Prior distributions of the NMPs. e$_0$ and $\rho_0$ are assumed as Gaussian distributions with a mean($\mu$) and standard deviation ($\sigma$). All other NMPs are assumed to be uniformly distributed between a minimum (min.) and a maximum (max.) value as listed in the units of MeV. In below cells, we have listed the prior distributions of CS model parameters. While $\rho_0$, $n_b$, $n_p$, and $w_p$ are in fm$^{-3}$ units, $h_p$ and $s_p$ are unitless. } 
  \centering
\setlength{\tabcolsep}{1.pt}
\renewcommand{\arraystretch}{1.4}
  \begin{ruledtabular}
  \begin{tabular}{ccccccccccc}
  %\toprule
&{$\rho_0$} & {e$_0$} &{K$_0$} & {Q$_0$} & {Z$_0$} &{J$_0$} & {L$_0$} &{K$_{\rm sym,0}$} & {Q$_{\rm sym,0}$}& {Z$_{\rm sym,0}$}\\ [1.3ex]
 \hline
 min./$\mu$ & 0.160 & -16.0 & 150 & -1500 & -5000 & 25 & 20 & -400 & -1500 &-5000\\[1.3ex] 
 max./$\sigma$ & 0.005 & 0.3 & 330 & 1500 & 5000 & 40 & 120 & 400 & 1500 &5000\\[1.3ex] 
 \hline \hline
 &{$n_b$} & {$h_p$} &{$n_p$} & {$w_p$} & {$s_p$} &  &  & & & \\ [1.3ex]
 \hline
 min. & 0.01 & 0.0 & 0.42 & 0.1 & -50 &  &  &  & &\\[1.3ex] 
 max. & 3.0 & 0.9 & 5.0 & 5.0 & 50 &  &  &  &  &\\[1.3ex] 
  \end{tabular}
  \end{ruledtabular}
\end{table}

\subsection{Speed-of-Sound (CS)}\label{CS}

We employ the speed-of-sound (CS) parameterization of Tews \textit{et al.}~\cite{Tews:2018kmu}. This assumption is consistent with hydrostatic equilibrium, which requires a monotonically increasing pressure. We adopt the transition density $\rho_{\rm tr} = 0.34\,\mathrm{fm}^{-3}$ to apply the $\chi$EFT constraint, ensuring consistency with other models for comparison. Beyond the  $\rho_{\rm tr}$, we impose causality ($c_s^2 \leq c^2$) and the conformal limit ($c_s^2/c^2 \to 1/3$). In practice, for $\rho < \rho_{\rm tr}$ we use the $n/3$ expansion [Eqs.~(\ref{n3E}, \ref{n3J})], while for $\rho > \rho_{\rm tr}$ the speed-of-sound model provides a smooth and causal extension of the EoS to high densities,

\bea
\frac{c_s^2}{c^2} &=& \frac{1}{3} - c_1 exp\left[{-\frac{(\rho-c_2)^2}{n_b^2}}\right] + h_p exp\left[-\frac{(\rho-n_p)^2}{w_p^2}\right]\nonumber\\
&& \left[1 + erf(s_p \frac{\rho-n_p}{w_p})\right], \label{eq-vs}
\eea 
where \(h_p\) determines the peak height or maximum value of the speed of sound, while the position \(n_p\) specifies the density at which it occurs. The width of the curve is controlled by the parameters \(w_p\) and \(n_b\), and the shape (or skewness) is set by the parameter \(s_p\). For a given value of \(n_b\), the coefficients \(c_1\) and \(c_2\) are fixed by requiring continuity of both the speed of sound and its derivative at the transition density \(\rho_{\rm tr}\). The parameters \(n_b\), \(h_p\), \(n_p\),  \(w_p\) and \(s_p\) are sampled from uniform distributions, as listed in Table~\ref{tab1:prior}.

We construct the high-density EoS beginning at the transition density \(\rho_{\rm tr}\), where the energy density \(\varepsilon(\rho_{\rm tr})\), the pressure \(P(\rho_{\rm tr})\), and the derivative of the energy density \(\varepsilon'(\rho_{\rm tr})\) are specified. The subsequent values of \(\varepsilon\) and \(P\) are then obtained iteratively using a step size of \(\Delta \rho = 0.001\,\mathrm{fm}^{-3}\), as follows:

\bea
 \rho_{i+1} &=& \rho_i + \Delta \rho \label{eq-rhoE},\\
 \varepsilon_{i+1} &=& \varepsilon_i + \Delta \varepsilon \nonumber\\
                &=& \varepsilon_i + \Delta \rho \frac{\varepsilon_i + P_i}{\rho_i}\label{eq-engE},\\
P_{i+1} &=&P_i + c_s^2(\rho_i) \Delta \varepsilon,  \label{eq-preE}.              
\eea
where the index i = 0 refers to the transition density $\rho_{\rm tr}$ . Note, in Eq.(\ref{eq-engE}) $\Delta e$ has been evaluated using the thermodynamic relation $ P =  \rho \partial \varepsilon/{\partial \rho}-\varepsilon$ valid at zero temperature.

For a selected model, $e(\rho,0)$ and $J(\rho)$ can be calculated using the relevant equations [e.g., Eqs.~(\ref{esnm_skyrme}) \& (\ref{esym_skyrme}) for Skyrme] for a given set of NMPs. These quantities are then used in Eq.~(\ref{eq:EoS}) to obtain the energy per particle and the corresponding energy density, $\varepsilon$. The neutron and proton chemical potentials are computed as $\mu_i = \partial \varepsilon / \partial \rho_i$, while the electron and muon chemical potentials follow from $\beta$-equilibrium, $\mu_n - \mu_p = \mu_e$ and $\mu_e = \mu_\mu$, together with charge neutrality, $\rho_p = \rho_e + \rho_\mu$, where $\rho_e$ and $\rho_\mu$ are the electron and muon number densities. The pressure is then obtained from the thermodynamic relation:
\begin{equation}
p = \sum_i \mu_i \rho_i - \varepsilon.
\end{equation}

To construct the whole EoS from the surface to the center of NS we generally divide the NS in three sections : outer crust,  inner crust and core. The outer crust EoS (up to $0.0016\rho_0$) follows the Baym-Pethick-Sutherland model~\cite{Baym:1971pw}, while the inner crust ($0.0016\rho_0 < \rho < \rho_{t1} \equiv 0.5\rho_0$) is modeled using a polytropic form~\cite{Carriere:2002bx, Patra:2022yqc}. Above $\rho_{t1}$, the core is constructed using the considered EoS models discussed above except for the CS model, which starts at $\rho_{t2}=0.34$ fm$^{-3}$, with the $n/3$ model bridging the intermediate region. Due to the use of the $n/3$ model in the intermediate region, the nuclear matter parameters come into play in the CS parameterization. All the valid EoSs should satisfy three essential conditions: (i) thermodynamic stability, (ii) causality ($c_s^2 < c^2$), and (iii) positive symmetry energy.

\section{Bayesian Analysis}\label{Bayes}

Bayesian likelihood is a central concept in Bayesian statistics, enabling the updating of probability estimates for a hypothesis as new data become available. This iterative updating underlies \textit{Bayesian inference}, where the aim is to refine our knowledge of the hypothesis parameters based on observed data. At its core is \textit{Bayes' theorem}, which relates prior information, observed data, and the hypothesis to yield the \textit{posterior probability} distribution over parameters. For a dataset $\mathcal{D}$ and a hypothesis $H(\theta)$, Bayes' theorem is expressed as

\[
P(\theta | \mathcal{D}, H) = \frac{{\mathcal{L}(\mathcal{D} | \theta, H) P(\theta | H)}}{P(\mathcal{D} | H)}
\]

The following is what each component represents:

\begin{itemize}
    \item $P(\theta | \mathcal{D}, H)$: the \textit{posterior probability} of parameters $\theta$ given data $\mathcal{D}$ under hypothesis $H$, representing our updated belief after observing the data.
    \item $\mathcal{L}(\mathcal{D} | \theta, H)$: the \textit{likelihood function}, describing the probability of observing $\mathcal{D}$ for a given set of parameters $\theta$ under $H$.
    \item $P(\theta | H)$: the \textit{prior probability} of $\theta$ under hypothesis $H$, reflecting our knowledge or assumptions before observing the data.
    \item $P(\mathcal{D} | H)$: the \textit{marginal likelihood} (or \textit{evidence}) under $H$, acting as a normalizing constant to ensure the posterior sums to 1 and accounting for the likelihood of the data over all parameter values.
\end{itemize}

\subsection{Likelihood} 

In Bayesian analysis, the likelihood function is defined as the probability of the observed data under a specific model, parametrized by $\theta$. It can be computed for a given set of experimental data and for posterior distributions from astrophysical observations as follows:

(i) \textbf{Experimental data:} for experimental data $D_{\rm expt} \pm \sigma$ with a symmetric Gaussian distribution, the likelihood is given by
\bea
{\mathcal L}({\mathcal D_{expt} | }\theta) &=& \frac{1}{\sqrt{2\pi\sigma^2}} exp(\frac{-(D(\theta)-D_{expt})^2}{2\sigma^2})\nonumber \\
&=& \mathcal{L}^{\rm expt}
\eea

Here \(D(\theta)\) is the model value for a given model parameter set \(\theta\). 

(ii) \textbf{GW observations:} for GW data, information on the EoS parameters is obtained from the component masses $m_1, m_2$ and the corresponding tidal deformabilities $\Lambda_1, \Lambda_2$. In this case,

\begin{align}
    P(d_{\mathrm{GW}}|\mathrm{EoS}) = \int^{M_u}_{m_2}dm_1 \int^{m_1}_{M_l} dm_2 P(m_1,m_2|\mathrm{EoS})   \nonumber \\
    \times P(d_{\mathrm{GW}} | m_1, m_2, \Lambda_1 (m_1,\mathrm{EoS}), \Lambda_2 (m_2,\mathrm{EoS})) \nonumber \\
    =\mathcal{L}^{\rm GW}
    \label{eq:GW-evidence}
\end{align}
where P(m$|$EoS)~\cite{Agathos:2015uaa, Wysocki:2020myz, Landry:2020vaw, Biswas:2020puz} can be written as 
\begin{equation}
    P(m|\rm{EoS}) = \left\{ \begin{matrix} \frac{1}{M_u - M_l} & \text{ iff } & M_l \leq m \leq M_u, \\ 0 & \text{ else, } & \end{matrix} \right.
\end{equation}
In our calculation we set $M_l$ = 1.36 M$_{\odot}$ and $M_u$ =1.6 M$_{\odot}$

(iii) \textbf{X-ray observations (NICER):} X-ray measurements provide the mass and radius of neutron stars. Accordingly, the corresponding likelihood is expressed as,

\begin{align}
    P(d_{\rm X-ray}|\mathrm{EoS}) = \int^{M_u}_{M_l} dm P(m|\mathrm{EoS}) \nonumber \\ \times
    P(d_{\rm X-ray} | m, R (m, \mathrm{EoS})) \nonumber \\
    = \mathcal{L}^{\rm NICER}
\end{align}
 \\
Here, $M_l$ represents a mass of 1 $M_\odot$, and $M_u$ denotes the maximum mass of a neutron star according to the respective EoS.\\

\vspace{0.5cm}
Suppose there are three independent datasets $A$, $B$, and $C$, with Bayesian likelihoods $\mathcal{L}^A(\theta)$, $\mathcal{L}^B(\theta)$, and $\mathcal{L}^C(\theta)$, where $\theta$ denotes the model parameters. When all these datasets are included in Bayesian inference, the total likelihood can be written as
\begin{equation}
   \mathcal{L}(\theta) = \prod_{i=A,B,C}\mathcal{L}^i(\theta) \nonumber 
\end{equation}

The final likelihood for the Set4 of our calculations(see Table~-\ref{tab2:data}) is \\

\begin{equation}
    \mathcal{L} = 
    \mathcal{L}^{\rm \chi EFT}
    \mathcal{L}^{\rm EXPT}
    \mathcal{L}^{\rm GW}
    \mathcal{L}^{\rm NICER I}
    \mathcal{L}^{\rm NICER II}\mathcal{L}^{\rm empirical}.
    \label{eq:lhd-set4}
\end{equation}
 
\begin{table}[t]
\caption{\label{tab2:data}Overview of constraints used in each scenario. Here \(\rho_c\)= 0.1 fm$^{-3}$ is the crossing density (see Sec.~\ref{EI}).}
\centering
\begin{ruledtabular}
\begin{tabular}{cc}
\textbf{Scenario} & \textbf{constraints} \\
\hline
Set1 & $\chi$EFT \\
Set2 & HIC,GW170817,NICER I, e$_{\mathrm{PNM}}(\rho_c)\&M(\rho_c)$ \\
Set3 & Set2 + NICER II\\
Set4 & Set1 + Set3 \\
Set5 & Set4 [Without e$_{\mathrm{PNM}}(\rho_c)\&M(\rho_c)$ ] \\
\end{tabular}
\end{ruledtabular}
\end{table}

\subsection{Bayes Factor}

The \textit{Bayes factor} is a ratio that compares the evidence for two hypotheses \( H_1 \) and \( H_2 \). It is defined as:
\bea
\text{BF}_{12} = \frac{P(D | H_1)}{P(D | H_2)}
\eea
where,
\begin{itemize}
    \item \( P(D | H_1) \) is the likelihood of the data \( D \) under hypothesis \( H_1 \).
    \item \( P(D | H_2) \) is the likelihood of the data \( D \) under hypothesis \( H_2 \).
\end{itemize}

\subsubsection{Log of the Bayes Factor }

The \textit{logarithm} of the Bayes factor simplifies the interpretation. It transforms the Bayes factor into a scale that is easier to work with and interpret. The log Bayes factor is:
\bea
\ln(\text{BF}_{12}) = \ln\left(\frac{P(D | H_1)}{P(D | H_2)}\right),\label{logBayes}
\eea
This gives you a more convenient way to express how strongly the data support one hypothesis over the other.
\subsubsection{Interpretation of the Bayes Factor }\label{IntpBF}
A common scale for interpreting the magnitude of the Bayes factor (or log Bayes factor) is as follows \cite{Kass01061995, Jarosz2014}:
\begin{itemize}
    \item \( \ln(\text{BF}) > 1 \): strong evidence for hypothesis \( H_1 \)
    \item \( 0 < \ln(\text{BF}) < 1 \): weak to moderate evidence for hypothesis \( H_1 \)
    \item \( \ln(\text{BF}) = 0 \): no evidence for either hypothesis
    \item \( -1 < \ln(\text{BF}) < 0 \): weak to moderate evidence for hypothesis \( H_2 \)
    \item \( \ln(\text{BF}) < -1 \): strong evidence for hypothesis \( H_2 \)
\end{itemize}

\section{Multiphysics Data}\label{data}
We employ five models—Taylor, $n/3$, Skyrme, RMF, and CS (Sec.~\ref{EoS})—to compute nuclear matter properties across densities and construct the corresponding neutron star EoSs. Below, we summarize the theoretical inputs, experimental and empirical inputs, and astrophysical observations used in this study.

\subsection{$\chi$EFT}
Using lattice QCD at finite density is difficult because of the sign problem in Monte Carlo simulations. Instead, at low densities, chiral effective theory works well with very small uncertainty. In particular, precise $N^{3}\mathrm{LO}$ calculations, where the underlying $NN+3N$ forces are constrained by nucleon--nucleon scattering phase shifts, few-body data (e.g., the triton), and the empirical nuclear saturation point, can be employed~\cite{Hebeler:2013nza, Drischler:2021kxf, Drischler:2020hwi}. In this work, we take the pure neutron matter pressure ($P_{\rm PNM}$) from N$^3$LO calculations up to 0.34 fm$^{-3}$~\cite{Drischler:2020hwi, Drischler:2021kxf}.

\subsection{Heavy Ion Collision}

We use empirical values of the symmetry energy $J(\rho)$, symmetry energy pressure $P_{sym}$, and symmetric nuclear matter pressure $P_{\rm SNM}$, extracted from experimental data on nuclear masses, neutron-skin thickness of $^{208}\mathrm{Pb}$, dipole polarizability, isobaric analog states, and heavy ion collisions over the density range 0.03–0.32 fm$^{-3}$, as summarized in Table~ I of Ref.~\cite{Tsang:2023vhh, Imam:2024gfh}.

\subsection{Empirical Input} \label{EI}
A notable feature observed across various energy density functionals (EDFs) is the convergence of their density-dependent incompressibility \( K(\rho) \) near the typical nuclear density, \( \rho_c \sim 0.1\, \text{fm}^{-3} \)~\cite{Khan_Jerome2013}. This behavior, stemming from EDFs calibrated to finite-nucleus properties, suggests that the slope of \( K(\rho) \) around \( \rho_c \), denoted \( \mathcal{M}_c \), may serve as a more relevant observable for constraining nuclear matter properties via giant monopole resonance (GMR) measurements than the incompressibility at saturation density \( K_0 \). Reference~\cite{Brown2013b} demonstrates that properties of doubly magic nuclei tightly constrain the neutron EoS at the crossing density \( \rho_c \sim 0.1\, \text{fm}^{-3} \).
 In our analysis, we utilize empirical determinations of \( \mathcal{M}_c \)~\cite{Khan_Jerome2013} and the energy per particle in pure neutron matter, \( e_{\rm PNM} \)~\cite{Brown2013b}, both evaluated at \( \rho_c \). These quantities are inferred from nuclear mass data and isoscalar GMR systematics. Importantly, \( e_{\rm PNM} \) reflects contributions from both the EoS of symmetric matter and the symmetry energy component.

\subsection{GW170817}

Gravitational wave observations of merging neutron stars with LIGO and Virgo allow us to measure their tidal deformability, which gives information about the dense-matter EoS. The GW170817 event, a binary neutron star merger with component masses ranging from 1.17 to 1.6~$M_\odot$, has led to many studies of neutron star properties~\cite{Abbott:2017vwq}. In this work, we use the publicly available posterior distributions of the tidal deformabilities from GW170817, provided by LIGO–Virgo Collaboration\footnote{LVK Collaboration,~\href{https://dcc.ligo.org/LIGO-P1800115/public}{https://dcc.ligo.org/LIGO-P1800115/public}}, as constraints on the EoS.

%\footnotetext[1]{LVK Collaboration, \href{https://dcc.ligo.org/LIGO-P1800115/public}{https://dcc.ligo.org/LIGO-P1800115/public}}

\subsection{NICER}
X-ray observations of surface hot spots on neutron stars with the NICER mission provide valuable constraints on the mass and radius of selected pulsars, both isolated and in binaries. These measurements, obtained by modeling the pulse profiles of rotating neutron stars, have become one of the most direct astrophysical probes of the dense-matter EoS. We organize the currently available results into two categories, NICER I and NICER II, as follows:

\subsubsection{NICER I}
The first set of constraints comes from the mass–radius posterior distributions of PSR J0030+0451 at its canonical mass, analyzed independently by two NICER groups: Riley {\sl et al.}~\cite{Riley:2019yda}~\footnote{~\href{https://zenodo.org/records/8239000}{https://zenodo.org/records/8239000}} and Miller {\sl et al.}~\cite{Miller:2019cac}~\footnote{~\href{https://zenodo.org/record/3473466}{https://zenodo.org/record/3473466}}. The second set of constraints comes from the mass–radius posterior distributions of PSR J0740+6620 at its maximum mass, similarly analyzed by Riley {\sl et al.}~\cite{Riley:2021pdl}~\footnote{~\href{https://zenodo.org/records/4697625}{https://zenodo.org/records/4697625}} and Miller {\sl et al.}~\cite{Miller:2021qha}~\footnote{~\href{https://zenodo.org/records/4670689}{https://zenodo.org/records/4670689}}. 

% Footnote texts
%\footnotetext[2]{\href{https://zenodo.org/records/8239000}{Riley PSR~J0030+0451 data}}
%\footnotetext[3]{\href{https://zenodo.org/record/3473466\#.XrOt1nWlxBc}{Miller PSR~J0030+0451 data}}
%\footnotetext[4]{\href{https://zenodo.org/records/4697625}{Riley PSR~J0740+6620 data}}
%\footnotetext[5]{\href{https://zenodo.org/records/4670689}{Miller PSR~J0740+6620 data}}

\subsubsection{NICER II}
More recently, NICER has provided additional mass–radius measurements for other pulsars, further expanding the dataset available for EoS studies. In particular, we use the publicly available posteriors for PSRJ0437+4715 from Choudhury {\sl et al.}~\cite{Choudhury:2024xbk}~\footnote{~\href{https://zenodo.org/records/13766753}{https://zenodo.org/records/13766753}} and PSRJ0614+3329 from Mauviard {\sl et al.}~\cite{Mauviard:2025dmd}~\footnote{~\href{https://zenodo.org/records/15603406}{https://zenodo.org/records/15603406}}. 
These new results mark an important step forward, as they extend NICER constraints beyond the initial two benchmark pulsars, providing broader coverage of the neutron star mass–radius plane.

%\footnotetext[6]{\href{https://zenodo.org/records/13766753}{Choudhury~ PSR~J0437+4715 data}}
%\footnotetext[7]{\href{https://zenodo.org/records/15603406}{Mauviard PSR~J0614+3329 data}}

\section{Results and Discussions}\label{PD}

We employ a wide range of EoS models—metamodels, nonrelativistic Skyrme-type mean field model, relativistic mean field model, and the speed-of-sound model—as detailed in Sec.~\ref{EoS}, to investigate the impact of diverse datasets on dense matter properties. These EoSs are used to determine neutron star properties such as mass, radius, and tidal deformability through the solutions of the Tolman–Oppenheimer–Volkoff equations~\cite{Oppenheimer:1939ne, Tolman:1939jz}. These models are further constrained within a Bayesian framework using experimental nuclear data, empirical nuclear inputs, theoretical predictions, and astrophysical observations. Table~\ref{tab1:prior} summarizes the prior distributions for model parameters, while Table~\ref{tab2:data} outlines the structured five-set data framework used to isolate the influence of individual constraints.

\subsection{Posterior distribution of EoS and NS Properties}\label{postD}

\begin{table*}[]
\caption{The 68$\%$ confidence interval distributions of the NMPs for all scenarios and all considered EoS models, with all NMPs expressed in MeV except $\rho_0$ which is in fm$^{-3}$. The symbols next to certain parameters (e.g. Q$_{\rm sym0}$) for Skyrme and RMF models indicate that these are not independent model parameters, so they are not computed.} \label{tab3:nmps}
\centering
\setlength{\tabcolsep}{1.pt}
\renewcommand{\arraystretch}{0.2}
\begin{ruledtabular}
\begin{tabular}{ccccccc}
{NMPs} & {Models} & Set1 & Set2 & Set3 & Set4 & Set5 \\

\cline{1-7}

& Taylor & 0.160$_{-0.005}^{+0.004}$ &0.166$_{-0.003}^{+0.004}$ &0.166$_{-0.004}^{+0.004}$ &0.167$_{-0.003}^{+0.003}$ &0.166$_{-0.004}^{+0.004}$\\[1.5ex]
& $n/3$ & 0.161$_{-0.004}^{+0.005}$ &0.165$_{-0.004}^{+0.004}$ &0.166$_{-0.004}^{+0.004}$ &0.166$_{-0.003}^{+0.004}$ &0.165$_{-0.004}^{+0.004}$\\[1.5ex]
{$\rho_0$}& Skyrme & 0.160$_{-0.004}^{+0.005}$ &0.164$_{-0.004}^{+0.004}$ &0.165$_{-0.004}^{+0.004}$ &0.166$_{-0.003}^{+0.003}$ &0.169$_{-0.004}^{+0.004}$\\[1.5ex]
& RMF & 0.160$_{-0.005}^{+0.005}$ &0.163$_{-0.004}^{+0.005}$ &0.165$_{-0.005}^{+0.004}$ &0.166$_{-0.003}^{+0.003}$ &0.162$_{-0.005}^{+0.004}$\\[1.5ex]
& CS & 0.161$_{-0.004}^{+0.005}$ &0.165$_{-0.004}^{+0.004}$ &0.165$_{-0.004}^{+0.004}$ &0.166$_{-0.003}^{+0.003}$ &0.162$_{-0.004}^{+0.004}$\\[1.5ex]
\hline

& Taylor & -15.98$_{-0.28}^{+0.28}$ &-15.82$_{-0.25}^{+0.24}$ &-15.86$_{-0.26}^{+0.24}$ &-15.84$_{-0.27}^{+0.26}$ &-16.02$_{-0.28}^{+0.28}$\\[1.5ex]
& $n/3$ & -15.99$_{-0.30}^{+0.28}$ &-15.85$_{-0.25}^{+0.27}$ &-15.85$_{-0.24}^{+0.25}$ &-15.82$_{-0.27}^{+0.26}$ &-15.99$_{-0.28}^{+0.30}$\\[1.5ex]
{e$_0$}& Skyrme & -16.06$_{-0.27}^{+0.26}$ &-15.81$_{-0.25}^{+0.28}$ &-15.81$_{-0.26}^{+0.25}$ &-15.71$_{-0.25}^{+0.25}$ &-15.74$_{-0.26}^{+0.26}$\\[1.5ex]
& RMF & -16.00$_{-0.28}^{+0.28}$ &-15.83$_{-0.29}^{+0.29}$ &-15.82$_{-0.28}^{+0.27}$ &-15.87$_{-0.24}^{+0.28}$ &-16.13$_{-0.30}^{+0.31}$\\[1.5ex]
& CS & -15.99$_{-0.30}^{+0.28}$ &-15.85$_{-0.23}^{+0.28}$ &-15.85$_{-0.22}^{+0.24}$ &-15.88$_{-0.22}^{+0.27}$ &-15.98$_{-0.28}^{+0.27}$\\[1.5ex]
\hline

& Taylor & 242$_{-58}^{+58}$ &295$_{-23}^{+19}$ &296$_{-23}^{+19}$ &293$_{-17}^{+16}$ &253$_{-24}^{+25}$\\[1.5ex]
& $n/3$ & 236$_{-58}^{+62}$ &283$_{-22}^{+21}$ &279$_{-23}^{+21}$ &281$_{-20}^{+19}$ &238$_{-26}^{+28}$\\[1.5ex]
{K$_0$}& Skyrme & 187$_{-25}^{+34}$ &257$_{-16}^{+15}$ &259$_{-16}^{+14}$ &265$_{-14}^{+12}$ &257$_{-20}^{+15}$\\[1.5ex]
& RMF & 285$_{-44}^{+31}$ &270$_{-30}^{+28}$ &273$_{-32}^{+32}$ &317$_{-24}^{+09}$ &215$_{-55}^{+53}$\\[1.5ex]
& CS & 236$_{-58}^{+62}$ &280$_{-23}^{+22}$ &282$_{-23}^{+22}$ &284$_{-19}^{+19}$ &267$_{-35}^{+31}$\\[1.5ex]
\hline

&  Taylor & 119$_{-827}^{+927}$ &-452$_{-88}^{+92}$ &-528$_{-89}^{+95}$ &-424$_{-65}^{+77}$ &-292$_{-92}^{+87}$\\[1.5ex]
& $n/3$ & -54$_{-939}^{+1017}$ &-756$_{-125}^{+161}$ &-810$_{-103}^{+145}$ &-541$_{-95}^{+102}$ &-405$_{-114}^{+121}$\\[1.5ex]
{Q$_0$}& Skyrme & -258$_{-317}^{+349}$ &-366$_{-48}^{+60}$ &-377$_{-43}^{+52}$ &-366$_{-43}^{+46}$ &-358$_{-48}^{+57}$\\[1.5ex]
& RMF & -1003$_{-182}^{+191}$ &-382$_{-82}^{+77}$ &-420$_{-65}^{+77}$ &-471$_{-94}^{+68}$ &-531$_{-84}^{+63}$\\[1.5ex]
& CS & -54$_{-939}^{+1017}$ &-745$_{-152}^{+170}$ &-804$_{-172}^{+178}$ &-861$_{-156}^{+172}$ &-735$_{-228}^{+209}$\\[1.5ex]
\hline

&Taylor & -3377$_{-1081}^{+1449}$ &417$_{-125}^{+115}$ &511$_{-136}^{+133}$ &344$_{-101}^{+90}$ &215$_{-105}^{+104}$\\[1.5ex]
& $n/3$ & -2207$_{-1831}^{+2062}$ &3593$_{-1207}^{+843}$ &3941$_{-1105}^{+688}$ &1906$_{-918}^{+711}$ &1742$_{-890}^{+702}$\\[1.5ex]
{Z$_0$}& Skyrme & -- &-- &-- &--&--\\[1.5ex]
& RMF & -- &-- &-- &--&--\\[1.5ex]
& CS & -2207$_{-1831}^{+2062}$ &2504$_{-2473}^{+1770}$ &2494$_{-2453}^{+1751}$ &2489$_{-1890}^{+1712}$ &1694$_{-2099}^{+2149}$\\[1.5ex]
\hline

&Taylor & 32.58$_{-4.89}^{+4.86}$ &34.33$_{-1.69}^{+1.92}$ &32.80$_{-1.34}^{+1.65}$ &33.63$_{-0.69}^{+0.61}$ &32.44$_{-0.94}^{+0.91}$\\[1.5ex]
& $n/3$ & 31.93$_{-4.48}^{+4.89}$ &34.69$_{-1.52}^{+1.76}$ &33.15$_{-1.30}^{+1.44}$ &33.58$_{-0.73}^{+0.68}$ &32.30$_{-1.02}^{+1.02}$\\[1.5ex]
{J$_0$}& Skyrme & 34.19$_{-4.92}^{+2.50}$ &35.49$_{-1.37}^{+1.45}$ &34.08$_{-1.39}^{+1.42}$ &33.68$_{-0.55}^{+0.53}$ &33.89$_{-0.75}^{+0.60}$\\[1.5ex]
& RMF & 34.62$_{-1.17}^{+1.10}$ &33.96$_{-1.42}^{+1.57}$ &33.07$_{-1.43}^{+1.33}$ &33.38$_{-0.61}^{+0.59}$ &31.34$_{-0.87}^{+1.20}$\\[1.5ex]
& CS & 31.93$_{-4.48}^{+4.89}$ &34.62$_{-1.34}^{+1.57}$ &33.25$_{-1.28}^{+1.18}$ &33.77$_{-0.56}^{+0.69}$ &32.48$_{-0.99}^{+0.96}$\\[1.5ex]
\hline

& Taylor & 53$_{-3}^{+3}$ &63$_{-17}^{+20}$ &44$_{-13}^{+16}$ &54$_{-2}^{+2}$ &53$_{-3}^{+3}$\\[1.5ex]
&$n/3$ & 55$_{-3}^{+3}$ &68$_{-15}^{+18}$ &49$_{-13}^{+14}$ &56$_{-3}^{+3}$ &55$_{-3}^{+3}$\\[1.5ex]
{L$_0$}& Skyrme & 52$_{-3}^{+3}$ &80$_{-11}^{+13}$ &65$_{-11}^{+11}$ &56$_{-2}^{+2}$ &58$_{-3}^{+2}$\\[1.5ex]
& RMF & 55$_{-3}^{+4}$ &58$_{-9}^{+13}$ &50$_{-8}^{+9}$ &55$_{-2}^{+2}$ &52$_{-3}^{+3}$\\[1.5ex]
& CS & 55$_{-3}^{+3}$ &69$_{-13}^{+14}$ &52$_{-12}^{+11}$ &56$_{-3}^{+3}$ &56$_{-3}^{+3}$\\[1.5ex]
\hline

&Taylor & -84$_{-57}^{+58}$ &-138$_{-85}^{+88}$ &-250$_{-50}^{+66}$ &-173$_{-17}^{+18}$ &-135$_{-25}^{+25}$\\[1.5ex]
&$n/3$ & -89$_{-63}^{+58}$ &-95$_{-94}^{+92}$ &-209$_{-74}^{+86}$ &-139$_{-19}^{+18}$ &-99$_{-26}^{+27}$\\[1.5ex]
{K$_{sym0}$}& Skyrme & -59$_{-37}^{+26}$ &-25$_{-55}^{+77}$ &-88$_{-37}^{+47}$ &-132$_{-14}^{+15}$ &-121$_{-16}^{+20}$\\[1.5ex]
& RMF & -133$_{-37}^{+42}$ &-141$_{-35}^{+33}$ &-148$_{-33}^{+45}$ &-187$_{-13}^{+38}$ &-82$_{-53}^{+59}$\\[1.5ex]
& CS & -89$_{-63}^{+58}$ &-87$_{-64}^{+76}$ &-163$_{-75}^{+80}$ &-142$_{-27}^{+22}$ &-135$_{-37}^{+38}$\\[1.5ex]
\hline

& Taylor & 224$_{-913}^{+840}$ &827$_{-653}^{+436}$ &682$_{-519}^{+464}$ &860$_{-113}^{+110}$ &723$_{-126}^{+118}$\\[1.5ex]
&$n/3$ & 25$_{-1005}^{+944}$ &981$_{-434}^{+328}$ &973$_{-384}^{+326}$ &749$_{-112}^{+119}$ &625$_{-124}^{+130}$\\[1.5ex]
{Q$_{sym0}$}& Skyrme & -- &-- &-- &--&--\\[1.5ex]
& RMF & -- &-- &-- &--&--\\[1.5ex]
& CS & 25$_{-1005}^{+944}$ &1111$_{-383}^{+267}$ &1151$_{-324}^{+237}$ &744$_{-177}^{+202}$ &650$_{-210}^{+218}$\\[1.5ex]
\hline

& Taylor & -3390$_{-1016}^{+1472}$ &806$_{-1307}^{+1307}$ &1450$_{-1163}^{+1069}$ &-723$_{-302}^{+421}$ &-589$_{-303}^{+377}$\\[1.5ex]
&$n/3$ & -1822$_{-1983}^{+1957}$ &457$_{-3261}^{+2807}$ &904$_{-2948}^{+2620}$ &-4148$_{-604}^{+926}$ &-4118$_{-594}^{+933}$\\[1.5ex]
{Z$_{sym0}$}& Skyrme & -- &-- &-- &--&--\\[1.5ex]
& RMF & -- &-- &-- &--&--\\[1.5ex]
& CS & -1822$_{-1983}^{+1957}$ &724$_{-3070}^{+2849}$ &730$_{-2952}^{+2591}$ &-1564$_{-2110}^{+2714}$ &-1487$_{-2342}^{+2780}$\\[1.5ex]
\end{tabular}
\end{ruledtabular}
\end{table*}

Table~\ref{tab3:nmps} presents the posterior distributions of NMPs for all scenarios. In \textbf{Set1}, only the pressure of pure neutron matter from $\chi$EFT~\cite{Drischler:2020hwi} is used. This tightly constrains the symmetry energy slope $L_0 \sim 52 \pm 3$ MeV, consistent across models due to the relation $L_0 = 3P_{\mathrm{PNM}}(\rho_0)/\rho_0$.  With $\rho_0 \sim 0.16$ fm$^{-3}$ and $P_{\mathrm{PNM}}(\rho_0) \sim $ 2.9 MeV.fm$^{-3}$, $L_0\approx 54$ MeV. However, this value is notably different from PREX-II~\cite{prex2} and CREX~\cite{crex}. While $J_0$ is well constrained within RMF models, $K_{\mathrm{sym0}}$ exhibits model dependence, being lower in RMF and higher in Skyrme. The incompressibility $K_0$ remains $\sim 240$~MeV in Taylor, $n/3$, and CS models, but shows inverse trends relative to $K_{\mathrm{sym0}}$ in Skyrme and RMF.  The remaining NMPs are largely unconstrained in Set1. 

\begin{figure}
    \centering
    \includegraphics[width=\linewidth]{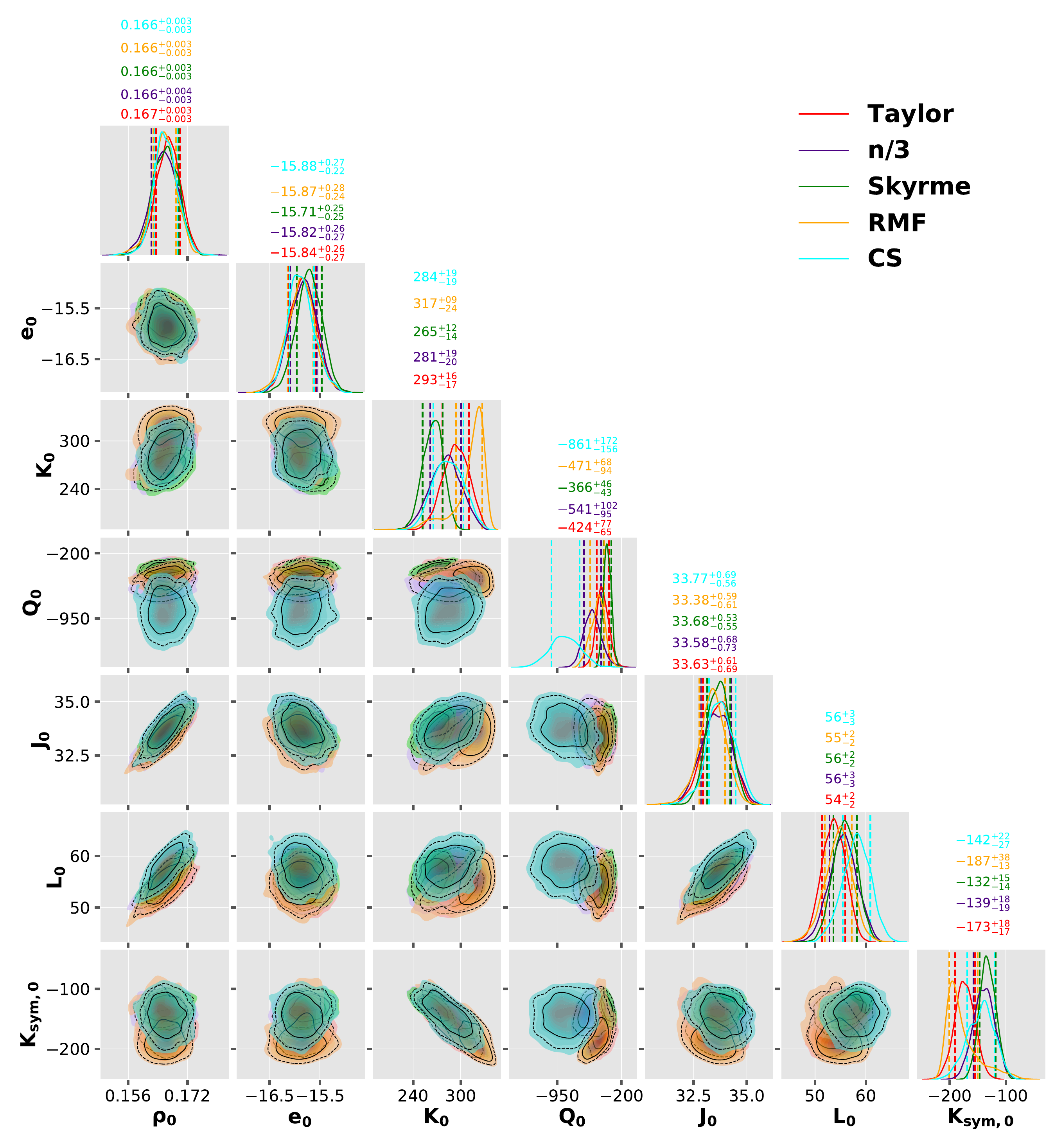}
    \caption{The marginalized posterior distributions of the NMPs, obtained through Bayesian inference using the Set4 scenario dataset, for all the considered EoS models with all NMPs expressed in MeV except $\rho_0$ which is in fm$^{-3}$.}
    \label{fig2}
\end{figure}

\vspace{2cm}
\begin{figure*}
   \centering 
 \includegraphics[width=14cm,height=10cm]{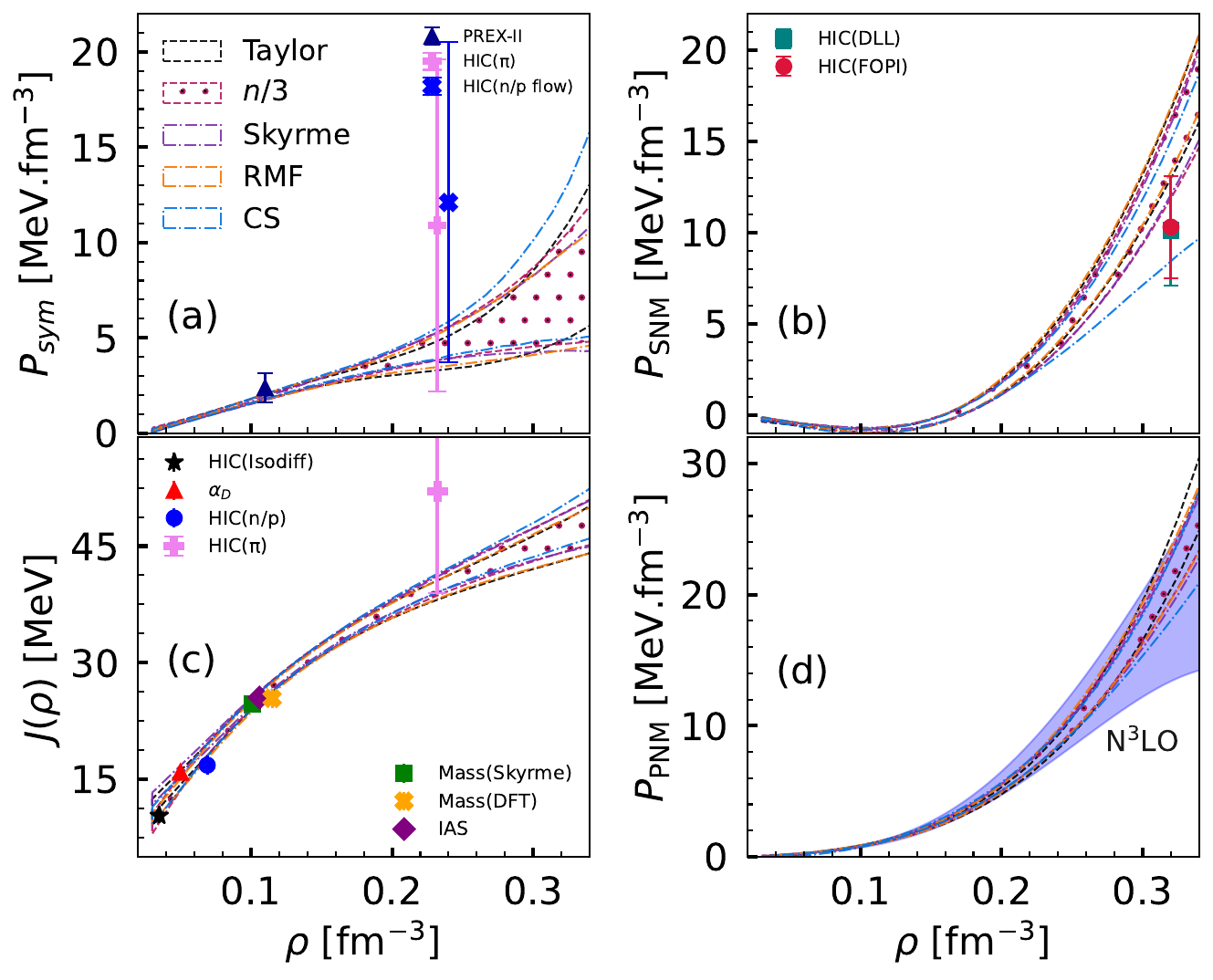}
    \caption{The 95\% confidence interval plots for (a) the pressure of the symmetry energy, $P_{\rm sym}$, (b) the pressure of SNM, $P_{\rm SNM}$, (c) the symmetry energy, $J(\rho)$, and (d) the pressure of PNM, $P_{\rm PNM}$, as a function of the baryon density $\rho$, evaluated using the posterior distributions from the Set4 scenario. The corresponding experimental data incorporated into the Bayesian framework are also shown.}
    \label{fig3}
\end{figure*}

\begin{figure*}{h!}
    \centering
  \includegraphics[width=16cm,height=6cm]{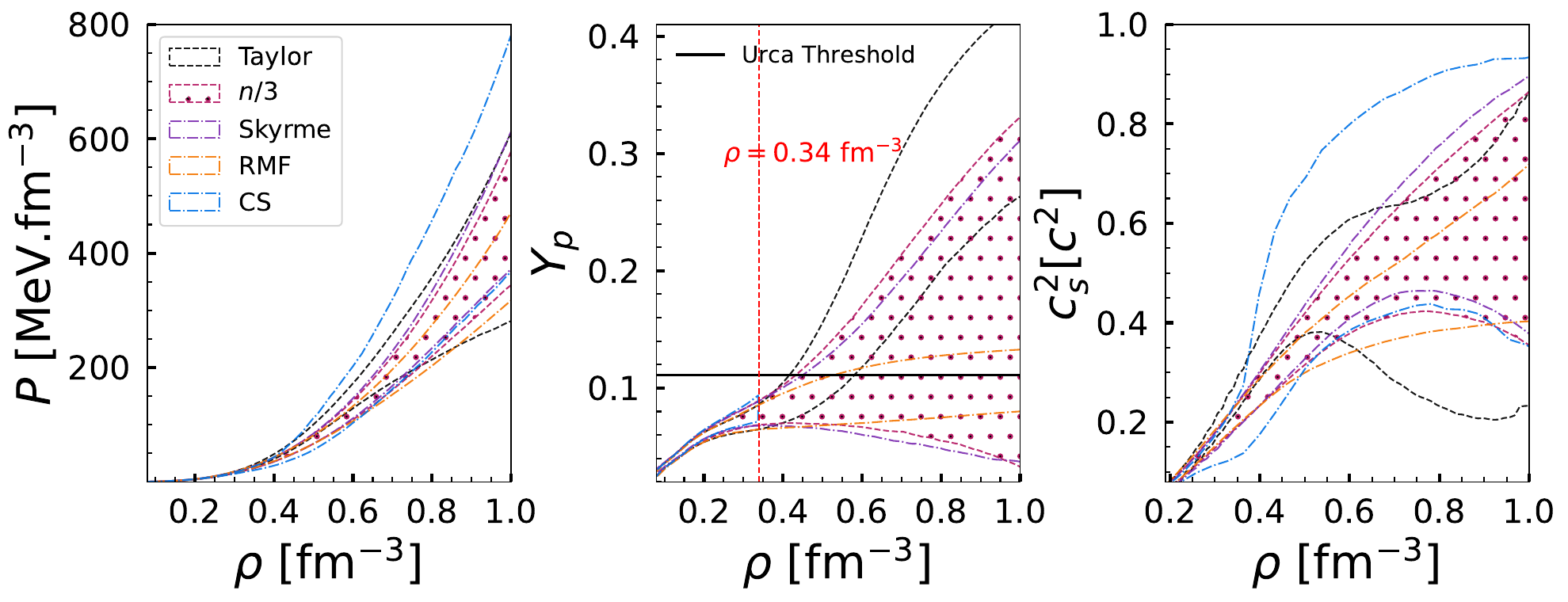}
    \caption{The 95$\%$ confidence interval posterior distributions of (a) the pressure of $\beta$-equilibrated matter, $P$, (b) the proton fraction, $Y_{\rm p}$, and (c) the squared sound speed, $c_s^2$, as functions of the baryon density $\rho$ obtained using the data in Set4 scenario. The onset of Urca cooling occurs above the solid black line. The vertical red dashed line indicates the transition density $\rho_{t2}=0.34$ fm$^{-3}$ for CS model.
}
    \label{fig4}
\end{figure*}

Notably, since the $n/3$ model is used in the sub-CS regime, its posterior distributions match those of the CS model. \textbf{Set2} incorporates heavy ion data, GW170817, NICER data for PSR J0030+0451 and PSR J0740+6620, and empirical constraints on energy per particle of PNM and slope of the incompressibility coefficients ${\mathcal M}$ at crossing density. These tighten the constraints on $K_0$, $Q_0$, and $J_0$ significantly, but increase uncertainties in $L_0$ and $K_{\mathrm{sym0}}$ due to the absence of a direct PNM constraint. In \textbf{Set3}, the inclusion of NICER-II data (PSR~J0437+4715 and PSR~J0614+3329) primarily affects symmetry energy parameters. $J_0$, $L_0$, and $K_{\mathrm{sym0}}$ decrease in median with reduced uncertainties—except in RMF and CS models. SNM parameters remain mostly unchanged. \textbf{Set4} combines \textbf{Set1}’s $\chi$EFT PNM data with Set3’s NICER-II constraints, leading to stiffer SNM with higher $K_0$ and $Q_0$ values across all models. Symmetry energy parameters are tightly constrained, with upward shifts in medians, except for Skyrme and RMF. In \textbf{Set5}, empirical constraints on $e_{\mathrm{PNM}}(\rho_c)$ and $M(\rho_c)$ are removed to resolve potential tension with $\chi$EFT PNM pressure. The analysis shows that $K_0$ decreases across all models, while $Q_0$ increases except in the RMF case. The parameter $J_0$ exhibits a reduction accompanied by larger uncertainties, whereas $L_0$ remains essentially unchanged. Both $K_{\mathrm{sym0}}$ and $Q_{\mathrm{sym0}}$ shift toward higher values with broader distributions. Bayesian evidence, particularly for the Skyrme model in Set4 ($\sim -98$) and Set5 ($\sim -89$), indicates that the empirical data are inconsistent with the $\chi$EFT PNM results. To reconcile this tension, the fits favor larger values of $K_0$ and smaller values of $Q_0$. Similar trends are observed for the other models.

Figure~\ref{fig2} shows the NMPs correlations for Set4, highlighting strong anticorrelation between $K_0$ and $K_{\mathrm{sym0}}$ ($r\simeq - 0.8$), and positive correlations of $J_0$ and $L_0$ is $r\simeq 0.8$ and also with $\rho_0$ is approximately $r\simeq 0.8$ for both cases. In Figure~\ref{fig3}, we plot 95 \% confidence interval for (a) the pressure of the symmetry energy, $P_{\rm sym}$, (b) the pressure of SNM, $P_{\rm SNM}$, (c) the symmetry energy, $J(\rho)$, and (d) the pressure of PNM, $P_{\rm PNM}$, as functions of the baryon density $\rho$, evaluated using the posterior distributions from the Set4 scenario.
All models agree reasonably with experimental data up to $2\rho_0$.
Figure~\ref{fig4} shows $P(\rho)$, $Y_p(\rho)$, and $c_s^2(\rho)$ for Set4. RMF shows the narrowest bounds, $c_s$ the widest. Up to $2\rho_0$, all models predict similar $Y_p$, diverging beyond. Taylor predicts the highest $Y_p$, potentially allowing for the direct Urca process, marked by the condition $k_F^p + k_F^e \ge k_F^n$. The squared speed of sound also shows significant variation, especially in the Taylor model, where $c_s^2$ drops beyond $\sim 3.5\rho_0$ (see Table~\ref{tab4:ns}).

\begin{figure*}
    \centering
    \includegraphics[width=\textwidth]{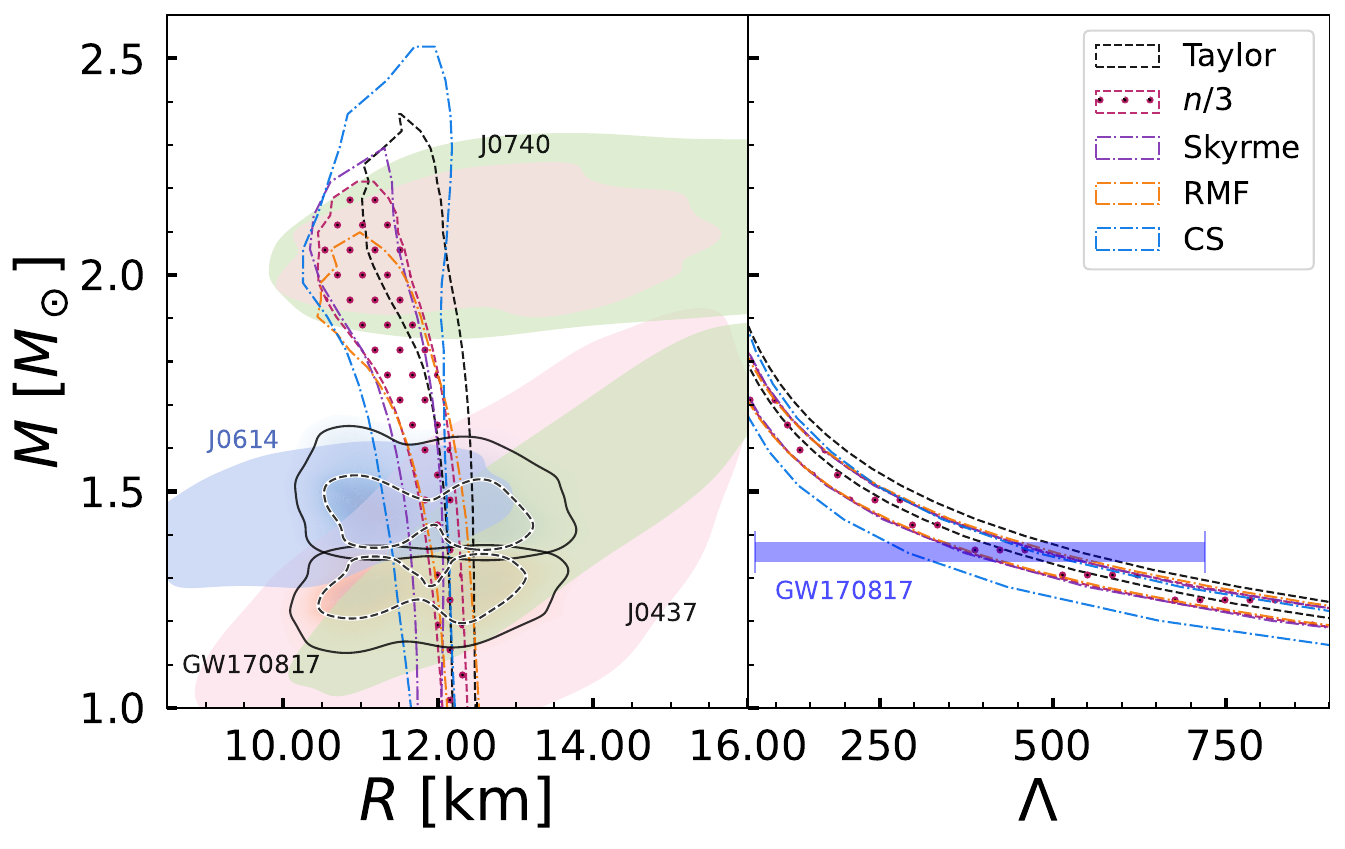}
    \caption{ The 95$\%$ confidence interval distributions for the radius, R(km) (left panel) and tidal deformability, $\Lambda$ (right panel) as a function of neutron star mass, $M(M_\odot)$, evaluated using the posterior distributions of the Set4 scenario. The astrophysical observations incorporated in the Bayesian framework are also shown.}
    \label{fig5}
\end{figure*}

Table~\ref{tab4:ns} summarizes the 68\% confidence intervals for NS properties across all valid EoSs (excluding Set1). In \textbf{Set2}, $M_{\rm max} \sim 2.10 \pm 0.08 M_\odot$ across models, slightly lower in RMF . NICER-II in \textbf{Set3} reduces $M_{\rm max}$, with further reduction in \textbf{Set4}. Removal of empirical inputs in \textbf{Set5} has minimal impact on $M_{\rm max}$. The radius $R_{1.4}$ for $1.4\,M_\odot$ NSs is $\sim 12.6 \pm 0.5$ km (Taylor, $n/3$), $\sim 12.7 \pm 0.35$ km (Skyrme, CS), and $\sim 12.5 \pm 0.35$ km (RMF). NICER-II in Set3 reduces $R_{1.4}$ by 1 km (Taylor, $n/3$), and by 0.5 km in others. $\chi$EFT PNM constraints in Set4 raise $R_{1.4}$ in Taylor, $n/3$, while reducing it in other models. Set5 lowers $R_{1.4}$ slightly for Taylor, $n/3$, while others remain stable. Since $\Lambda_{1.4} \propto R^5$, its trend follows $R_{1.4}$. Values shift from $\sim 546 \pm 90$ (Set2) to $\sim 429 \pm 75$ (Set3) and $\sim 354 \pm 25$ (Set4). Set5 reduces it slightly further. Central densities $\rho_{c,1.4}$ increase with added NICER constraints—from $\sim 2.5 \rho_0$ in Set2 to $\sim 3 \rho_0$ in Set3 and beyond. Central pressure and squared sound speed follow similar behavior. The squared sound speed at the center of $1.4\,M_\odot$ stars increases in Set3 and then decreases in Set4 for the Taylor model, and shows similar trends across other models. $c_{s,\max}^2$ remains largely unaffected, except for RMF, which consistently shows the lowest values. Figure~\ref{fig5} displays the 95 \% confidence intervals for M–R and M–$\Lambda$ distributions for Set4. The CS model shows the broadest distributions of radii, while RMF exhibits the narrowest distributions at higher mass, and Taylor shows narrowest distributions at low mass. The CS model exhibits comparatively broader distributions for tidal deformability, while the other models show similar distributions.

\begin{figure*}
    \centering
    \includegraphics[width=\textwidth]{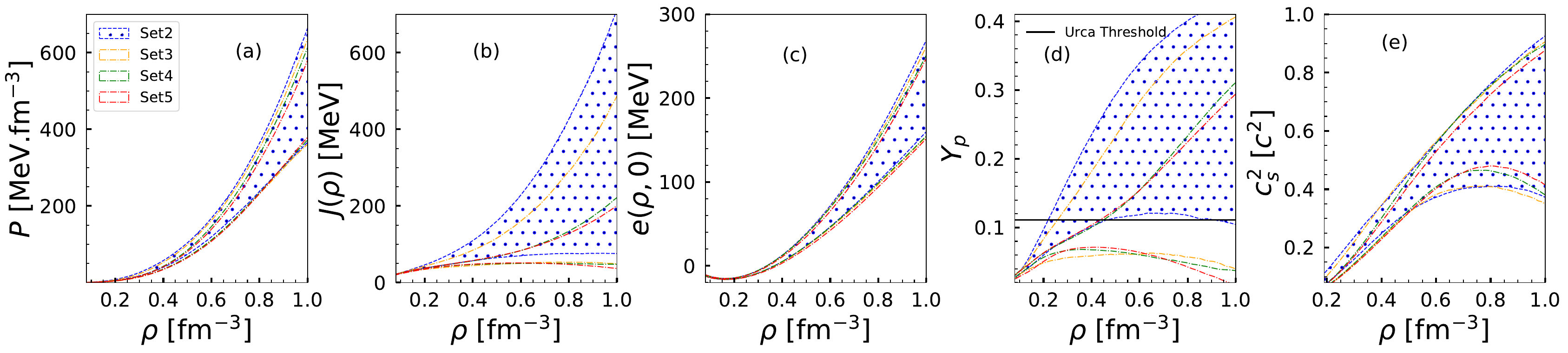}
    \caption{The 95$\%$ confidence interval posterior distributions of (a) the pressure of $\beta$-equilibrated matter, $P$, (b) symmetry energy,$J (\rho)$, (c) energy for SNM, $e(\rho,0)$  (d) the proton fraction, $Y_{\rm p}$, and (e) the squared sound speed, $c_s^2$, as a function of the baryon density $\rho$ obtained using the data for all sets in the Skyrme model.
    The onset of Urca cooling occurs above the solid black line.
}
    \label{fig6}
\end{figure*}
\begin{table*}[]
\caption{The 68\% confidence interval distributions for various NS properties and key EoS quantities, shown across all scenarios and for all considered EoS models.}\label{tab4:ns}
\centering
\setlength{\tabcolsep}{1.pt}
\renewcommand{\arraystretch}{0.7}
\begin{ruledtabular}
\begin{tabular}{cccccc}
{NS Properties} & {Models} & Set2 & Set3 & Set4 & Set5 \\
\cline{1-6}
& Taylor & 2.10$_{-0.08}^{+0.09}$ & 2.08$_{-0.08}^{+0.09}$ & 2.05$_{-0.08}^{+0.08}$ & 2.07$_{-0.08}^{+0.09}$\\[1.5ex]
& $n/3$ & 2.09$_{-0.08}^{+0.08}$ & 2.09$_{-0.08}^{+0.08}$ & 2.01$_{-0.07}^{+0.07}$ & 2.02$_{-0.07}^{+0.07}$\\[1.5ex]
{M$_{max}$(M$_\odot$)}& Skyrme & 2.09$_{-0.08}^{+0.08}$ & 2.06$_{-0.08}^{+0.08}$ & 2.06$_{-0.07}^{+0.06}$ & 2.04$_{-0.06}^{+0.05}$\\[1.5ex]
& RMF & 2.01$_{-0.08}^{+0.09}$ & 1.98$_{-0.09}^{+0.07}$ & 1.92$_{-0.05}^{+0.07}$ & 1.90$_{-0.05}^{+0.08}$\\[1.5ex]
& CS & 2.10$_{-0.10}^{+0.10}$ & 2.09$_{-0.10}^{+0.10}$ & 2.07$_{-0.08}^{+0.14}$ & 2.09$_{-0.10}^{+0.15}$\\[1.5ex]
\hline

& Taylor & 12.60$_{-0.63}^{+0.54}$ & 11.56$_{-0.38}^{+0.56}$ & 12.30$_{-0.09}^{+0.08}$ & 12.00$_{-0.10}^{+0.10}$\\[1.5ex]
& $n/3$ & 12.60$_{-0.51}^{+0.46}$ & 11.74$_{-0.52}^{+0.55}$ & 12.06$_{-0.11}^{+0.11}$ & 11.90$_{-0.13}^{+0.12}$\\[1.5ex]
{R$_{1.4}$(km)}& Skyrme & 12.70$_{-0.32}^{+0.35}$ & 12.22$_{-0.36}^{+0.30}$ & 11.85$_{-0.11}^{+0.11}$ & 11.77$_{-0.12}^{+0.11}$\\[1.5ex]
& RMF & 12.51$_{-0.36}^{+0.37}$ & 12.20$_{-0.33}^{+0.27}$ & 12.14$_{-0.14}^{+0.14}$ & 12.16$_{-0.09}^{+0.11}$\\[1.5ex]
& CS & 12.69$_{-0.36}^{+0.43}$ & 12.02$_{-0.47}^{+0.44}$ & 11.80$_{-0.21}^{+0.17}$ & 11.81$_{-0.22}^{+0.19}$\\[1.5ex]
\hline

& Taylor & 484$_{-130}^{+121}$ & 295$_{-58}^{+108}$ & 412$_{-23}^{+21}$ & 402$_{-24}^{+25}$\\[1.5ex]
& $n/3$ & 507$_{-105}^{+103}$ & 337$_{-91}^{+118}$ & 358$_{-25}^{+26}$ & 349$_{-26}^{+27}$\\[1.5ex]
{$\Lambda_{1.4}$}& Skyrme & 546$_{-88}^{+112}$ & 429$_{-73}^{+81}$ & 354$_{-26}^{+25}$ & 334$_{-25}^{+26}$\\[1.5ex]
& RMF & 410$_{-50}^{+57}$ & 375$_{-54}^{+51}$ & 365$_{-28}^{+28}$ & 367$_{-23}^{+24}$\\[1.5ex]
& CS & 509$_{-75}^{+89}$ & 387$_{-90}^{+93}$ & 324$_{-43}^{+37}$ & 322$_{-47}^{+39}$\\[1.5ex]
\hline

& Taylor & 0.42$_{-0.03}^{+0.05}$ & 0.49$_{-0.05}^{+0.05}$ & 0.45$_{-0.01}^{+0.02}$ & 0.46$_{-0.02}^{+0.02}$\\[1.5ex]
& $n/3$ & 0.42$_{-0.03}^{+0.04}$ & 0.48$_{-0.05}^{+0.06}$ & 0.50$_{-0.02}^{+0.02}$ & 0.51$_{-0.02}^{+0.02}$\\[1.5ex]
{$\rho_{c,1.4}$(fm$^{-3}$)}& Skyrme & 0.42$_{-0.03}^{+0.04}$ & 0.47$_{-0.04}^{+0.04}$ & 0.50$_{-0.02}^{+0.02}$ & 0.52$_{-0.02}^{+0.02}$\\[1.5ex]
& RMF & 0.47$_{-0.03}^{+0.03}$ & 0.49$_{-0.03}^{+0.04}$ & 0.51$_{-0.02}^{+0.02}$ & 0.50$_{-0.02}^{+0.02}$\\[1.5ex]
& CS & 0.43$_{-0.03}^{+0.03}$ & 0.47$_{-0.04}^{+0.05}$ & 0.52$_{-0.04}^{+0.04}$ & 0.52$_{-0.03}^{+0.04}$\\[1.5ex]
\hline

& Taylor & 59$_{-8}^{+14}$ & 81$_{-15}^{+14}$ & 66$_{-3}^{+3}$ & 68$_{-3}^{+3}$\\[1.5ex]
& $n/3$ & 58$_{-7}^{+10}$ & 75$_{-13}^{+17}$ & 75$_{-4}^{+4}$ & 77$_{-4}^{+5}$\\[1.5ex]
{$P_{c,1.4}$(MeV.fm$^{-3}$)}& Skyrme & 57$_{-7}^{+8}$ & 67$_{-8}^{+10}$ & 76$_{-4}^{+5}$ & 79$_{-4}^{+5}$\\[1.5ex]
& RMF & 68$_{-6}^{+7}$ & 72$_{-7}^{+9}$ & 75$_{-5}^{+5}$ & 74$_{-4}^{+4}$\\[1.5ex]
& CS & 58$_{-7}^{+7}$ & 70$_{-9}^{+14}$ & 81$_{-7}^{+9}$ & 81$_{-7}^{+11}$\\[1.5ex]
\hline

& Taylor & 0.39$_{-0.06}^{+0.12}$ & 0.59$_{-0.14}^{+0.14}$ & 0.39$_{-0.02}^{+0.02}$ & 0.39$_{-0.02}^{+0.02}$\\[1.5ex]
& $n/3$ & 0.35$_{-0.04}^{+0.07}$ & 0.46$_{-0.09}^{+0.12}$ & 0.36$_{-0.01}^{+0.01}$ & 0.36$_{-0.01}^{+0.01}$\\[1.5ex]
{$c_{s,1.4}^2$(c$^2$)}& Skyrme & 0.32$_{-0.03}^{+0.02}$ & 0.35$_{-0.02}^{+0.03}$ & 0.38$_{-0.01}^{+0.01}$ & 0.38$_{-0.01}^{+0.01}$\\[1.5ex]
& RMF & 0.35$_{-0.03}^{+0.02}$ & 0.35$_{-0.03}^{+0.02}$ & 0.33$_{-0.02}^{+0.01}$ & 0.34$_{-0.01}^{+0.02}$\\[1.5ex]
& CS & 0.33$_{-0.03}^{+0.04}$ & 0.38$_{-0.05}^{+0.07}$ & 0.42$_{-0.06}^{+0.12}$ & 0.46$_{-0.08}^{+0.11}$\\[1.5ex]
\hline
& Taylor & 0.95$_{-0.05}^{+0.06}$ & 0.97$_{-0.06}^{+0.07}$ & 1.01$_{-0.05}^{+0.05}$ & 1.04$_{-0.05}^{+0.05}$\\[1.5ex]
& $n/3$ & 1.03$_{-0.06}^{+0.06}$ & 1.06$_{-0.07}^{+0.07}$ & 1.14$_{-0.04}^{+0.06}$ & 1.18$_{-0.06}^{+0.03}$\\[1.5ex]
{$\rho_{c,\max}$(fm$^{-3}$)}& Skyrme & 1.05$_{-0.06}^{+0.07}$ & 1.11$_{-0.08}^{+0.08}$ & 1.13$_{-0.04}^{+0.07}$ & 1.19$_{-0.07}^{+0.02}$\\[1.5ex]
& RMF & 1.09$_{-0.07}^{+0.05}$ & 1.10$_{-0.02}^{+0.09}$ & 1.18$_{-0.09}^{+0.01}$ & 1.18$_{-0.09}^{+0.00}$\\[1.5ex]
& CS & 1.04$_{-0.06}^{+0.08}$ & 1.09$_{-0.08}^{+0.09}$ & 1.13$_{-0.10}^{+0.08}$ & 1.12$_{-0.10}^{+0.08}$\\[1.5ex]
\hline

& Taylor & 403$_{-99}^{+102}$ & 368$_{-80}^{+103}$ & 441$_{-104}^{+116}$ & 500$_{-116}^{+115}$\\[1.5ex]
& $n/3$ & 541$_{-112}^{+96}$ & 543$_{-134}^{+133}$ & 622$_{-127}^{+116}$ & 679$_{-122}^{+107}$\\[1.5ex]
{$P_{c,\max}$(MeV.fm$^{-3}$)}& Skyrme & 574$_{-120}^{+123}$ & 617$_{-146}^{+128}$ & 659$_{-137}^{+126}$ & 744$_{-154}^{+91}$\\[1.5ex]
& RMF & 507$_{-90}^{+110}$ & 520$_{-79}^{+115}$ & 520$_{-79}^{+115}$ & 456$_{-24}^{+42}$\\[1.5ex]
& CS & 568$_{-132}^{+136}$ & 639$_{-146}^{+124}$ & 705$_{-142}^{+127}$ & 700$_{-147}^{+122}$\\[1.5ex]
\hline

& Taylor & 0.62$_{-0.22}^{+0.22}$ & 0.56$_{-0.21}^{+0.22}$ & 0.59$_{-0.24}^{+0.25}$ & 0.64$_{-0.23}^{+0.22}$\\[1.5ex]
& $n/3$ & 0.75$_{-0.20}^{+0.13}$ & 0.70$_{-0.23}^{+0.17}$ & 0.66$_{-0.22}^{+0.19}$ & 0.73$_{-0.22}^{+0.17}$\\[1.5ex]
{$c_{s,\max}^2$(c$^2$)}& Skyrme & 0.72$_{-0.19}^{+0.16}$ & 0.70$_{-0.22}^{+0.18}$ & 0.68$_{-0.24}^{+0.22}$ & 0.81$_{-0.25}^{+0.14}$\\[1.5ex]
& RMF & 0.53$_{-0.09}^{+0.16}$ & 0.54$_{-0.10}^{+0.13}$ & 0.52$_{-0.09}^{+0.17}$ & 0.44$_{-0.03}^{+0.04}$\\[1.5ex]
& CS & 0.70$_{-0.23}^{+0.18}$ & 0.74$_{-0.21}^{+0.17}$ & 0.76$_{-0.19}^{+0.16}$ & 0.76$_{-0.23}^{+0.15}$\\[1.5ex]
\end{tabular}
\end{ruledtabular}
\end{table*}
\pagebreak
\begin{figure*}
    %\centering   
    \vspace{-0.25cm}
    \includegraphics[width=0.8\textwidth]{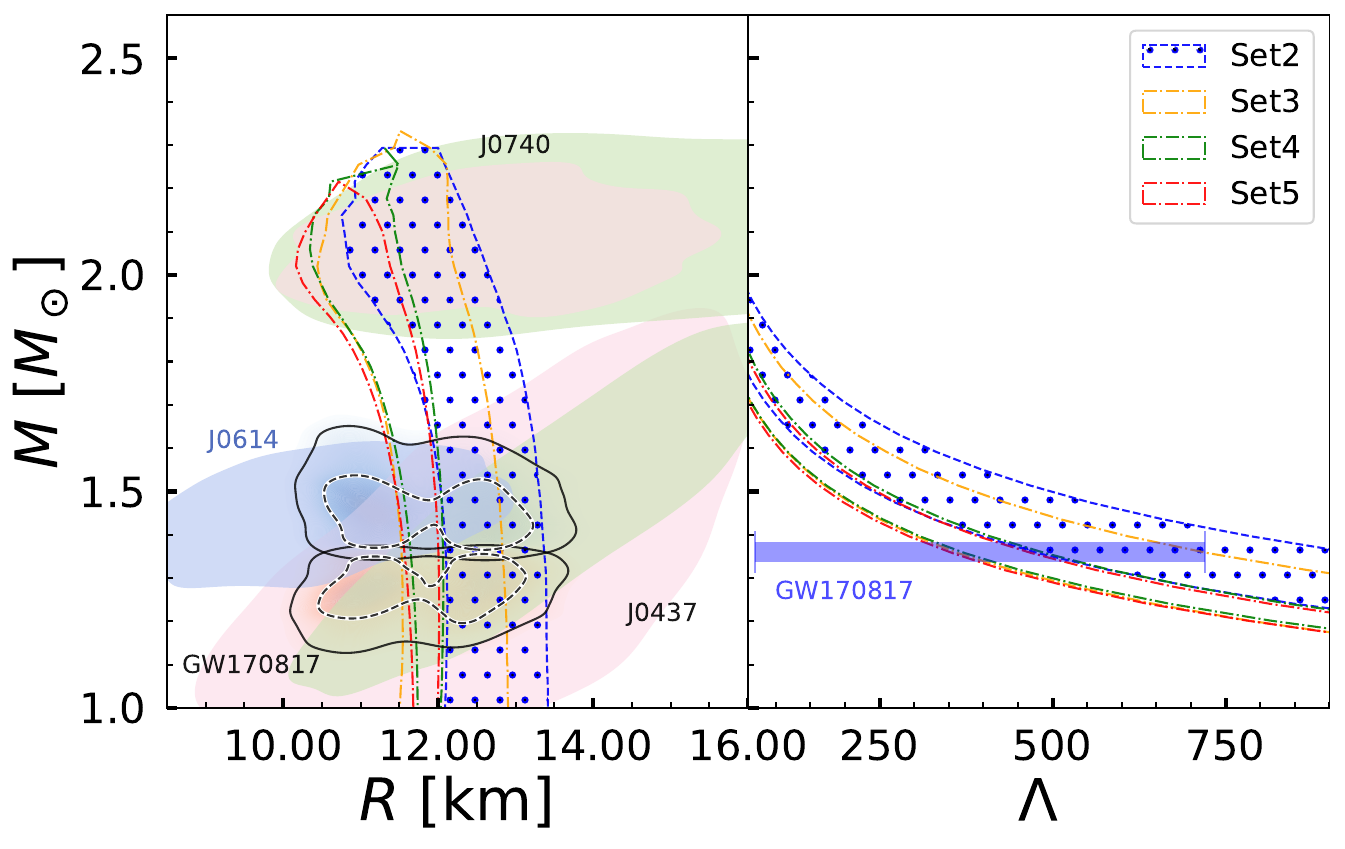}
    \caption{ The 95$\%$ confidence interval distributions for the radius, R(km) (left panel) and tidal deformability, $\Lambda$ (right panel) as functions of neutron star mass, $M(M_\odot)$, evaluated using the posterior distributions of all sets in case of the Skyrme model. The astrophysical observations incorporated in the Bayesian framework are also shown.}
    \label{fig7}
\end{figure*}

\subsection{Model Comparison}

Since we use a large number of EoS models, it would be good to compare these models statistically using the Bayes factor. In Table~\ref{tab5}, we present the logarithmic value of the Bayes factor computed using Eq.(\ref{logBayes}) for the Set4 scenario. The interpretation of the Bayes factor is discussed in Sec.~\ref{IntpBF}. We consider the Skyrme model as hypothesis H2 and other models as H1 to compute \(\ln(\text{BF}_{12})\). From the large negative values of \(\ln(\text{BF}_{12})\), it is clear that Skyrme is definitely the best model for the Set4 data in the Bayesian inference. We have also computed Bayes factor for all other scenarios and found the trend is more or less the same i.e. Skyrme being the best model but in Set2 and Set3 the CS model is better marginally with \(\ln(\text{BF}_{12})\)$\sim$ + 1.4 and +2.6, respectively.

\begin{table}[ht]
\caption{The logarithmic Bayes factor for all the EoS models considered for the Set4 scenario. Here we consider Skyrme as H2 hypothesis always and other models as H1 to compute \(\ln(\text{BF}_{12})\) using Eq. (\ref{logBayes}).\label{tab5}}
\centering
\begin{ruledtabular}
\begin{tabular}{cc}
\textbf{Model} & \textbf{\(\ln(\text{BF}_{12})\) } \\
\hline
Taylor & -14.53 \\
$n/3$ & -2.8\\
RMF &  -23.19\\
CS &  -19.26\\
\end{tabular}
\end{ruledtabular}
\end{table}

\subsection{Posterior distribution of the EoS and NS properties for the overall best model}

Given that the Skyrme model is favored by Bayes factor analysis, we examine how the distributions of EoS quantities and neutron star properties evolve across different datasets. In Fig.~\ref{fig6}, we present the 95\% confidence interval distributions of (a) pressure of $\beta$-equilibrated matter, $P$, (b) symmetry energy, $J(\rho)$, (c) energy per particle of symmetric nuclear matter, $e(\rho,0)$, (d) proton fraction, $Y_p$, and (e) squared speed of sound, $c_s^2$, as functions of baryon density $\rho$.

The addition of NICER measurements from PSR~J0614+3329 and PSR~J0437+4715 in Set3 significantly constrains these quantities, especially the symmetry energy. For instance, the symmetry energy at $2\rho_0$ and $4\rho_0$ changes from $63^{+9}_{-7}$ MeV and $142^{+62}_{-43}$~MeV in Set2 to $54^{+7}_{-7}$ MeV and $100^{+39}_{-29}$~MeV in Set3. When $\chi$EFT constraints on PNM are added in Set4, the bounds become much tighter: $47^{+2}_{-2}$ MeV and $73^{+13}_{-12}$~MeV, respectively. This tightening is also reflected in the distribution of $Y_p$ and $c_s^2$, where Set4 and Set5 show particularly narrow bands.

Figure~\ref{fig7} shows the mass-radius (M–R) and mass–tidal deformability (M–$\Lambda$) relations for the Skyrme model across all scenarios. The NICER radius measurements in Set3 constrain the radius of a $1.4 M_\odot$ NS to $R_{1.4} \sim 12.22 \pm 0.3$~km, narrower than in Set2. The inclusion of PNM constraints in Set4 reduces this further to a very tight bound $R_{1.4}$ = 11.85 $\pm$ 0.11 km, while the removal of empirical nuclear inputs in Set5 has negligible impact. This value is close to that obtained in Ref.~\cite{Ng:2025wdj}.
A similar trend is observed in the tidal deformability. The constraint from Set2 gives $\Lambda_{1.4} \sim 546 \pm 90$, which reduces to $\sim 429 \pm 75$ in Set3 with new NICER data. In Set4, with all constraints combined, it tightens further to $\Lambda_{1.4} \sim 354 \pm 25$. The absence of empirical data in Set5 slightly lowers this value but does not significantly affect the overall distribution.

These results underscore the impact of recent astrophysical observations—particularly NICER—combined with theoretical and experimental nuclear physics inputs, in tightly constraining the high-density behavior of the EoS.

\section{Conclusions} \label{Conclusion}
We employed Bayesian inference with five EoS models Taylor, $n/3$, Skyrme, RMF, and CS by integrating multiphysics constraints from nuclear experiments, empirical nuclear inputs, $\chi$EFT calculations of pure neutron matter, and neutron star observations to tightly constrain the nuclear matter parameters governing the EoS. The data were organized into five sets: \textbf{Set1} ($\chi$EFT PNM), \textbf{Set2} (terrestrial, empirical, and earlier astrophysical), \textbf{Set3} (new NICER radii of PSR~J0437+4715 and PSR~J0614+3329), \textbf{Set4} (all data combined), and \textbf{Set5} (excluding empirical nuclear inputs).

The PNM data from $\chi$EFT constrain the slope of the symmetry energy $L_0 \simeq 52 \pm 3$~MeV. Data in \textbf{Set2} significantly improve constraints on $K_0$, $Q_0$, and $J_0$, but broaden the uncertainties in $L_0$ and $K_{\mathrm{sym0}}$. The inclusion of recent NICER measurements for PSR~J0437+4715 and PSR~J0614+3329 further tightens the constraints on the symmetry energy and its density dependence, particularly $L_0$ and $K_{\mathrm{sym0}}$. The combined dataset in \textbf{Set4} yields the most stringent bounds on the nuclear matter parameters and the governing EoS.

Model comparisons revealed that the Skyrme interaction provides the best overall fit in the most data-rich scenarios (Sets 4 and 5), outperforming other models. RMF consistently yields lower maximum NS masses and narrower bounds on proton fraction and sound speed. 
The combined data constrain the key NMPs within narrow bounds. For the Skyrme model: $L_0 = 56 \pm 3$ MeV, $K_{\mathrm{sym}0}= -132 \pm 15$ MeV, $K_0 = 265 \pm 12$ MeV and $Q_0 = -366 \pm 43$ MeV.
Our results also yield precise NS observables:
$R_{1.4} = 11.85 \pm 0.11$ km and $\Lambda_{1.4} = 354 \pm 25$ 
with corresponding central densities of $\sim 3\rho_0 \pm 0.2\rho_0$ for a $1.4\,M_\odot$ star and $\sim 6$--$7\rho_0$ for the maximum mass star. 

This study underscores the power of synergistic constraints from theory, experiment, and astrophysical observations in pinning down the high-density EoS. Continued and more precise NS measurements will further refine these constraints.

\section{Author Contributions}

Both the authors contributed equally to this work, including the conceptualization of the problem, development of methodology, analysis, and preparation of the final manuscript.

\section{Acknowledgements}
The authors are grateful to Prof.\ Christian Drischler for providing the $\chi$EFT data and for valuable suggestions, and to Prof.\ Kai Zhou for his careful reading and insightful suggestions. We also thank Mr. Anagh Venneti for technical discussions.
A.I.\ acknowledges the use of the NLHPC (Centro de Modelamiento Matemático, Chile) for computational resources. N.K.P.\ acknowledges support from the CUHK–Shenzhen University Development Fund under Grant No.\ UDF01003041 and No. UDF03003041, as well as from the Shenzhen Peacock Fund under Grant No.\ 2023TC0007. Finally, the authors acknowledge the use of the analysis software BILBY~\cite{Ashton:2018jfp, Romero-Shaw:2020owr} and open data from GWOSC~\cite{Abbott2021}.

\section{Conflicts of Interest} The authors declare no conflicts of interest.
%\clearpage
%\newpage
 \bibliographystyle{apsrev4-1}
  \bibliography{main}

\end{document}